\documentclass[10pt,%
 amsmath,amssymb,
 aps,pre,twocolumn,
floatfix,
]{revtex4-1}
\usepackage[utf8]{inputenc}
\usepackage{amstext,amssymb,amsfonts,bbm,amsthm,bm}
\usepackage{mathtools}
\usepackage{hyperref}
\usepackage{graphicx}
\usepackage{subcaption}
\usepackage{caption}
\usepackage{booktabs} 
\newcommand{\pert}{\mathrm{pert}}
\newcommand{\unpert}{\mathrm{unper}}
\newcommand{\av}[1]{\overline{ #1}}

\newcommand{\ER}{Erd\H{o}s-R\'enyi }
\newcommand{\BA}{Barab\'asi-Albert }
\newcommand{\WS}{Watts–Strogatz }
\usepackage{xr} 
\usepackage{placeins} 

\begin{document}

\title{Exponential rate of epidemic spreading on complex networks}

\author{Samuel Cure}
\author{Florian G. Pflug}
\author{Simone Pigolotti}
\email{simone.pigolotti@oist.jp}
\affiliation{
Biological Complexity Unit, Okinawa Institute of Science and Technology, Onna, Okinawa 904-0495, Japan.
}

\date{\today}

\begin{abstract}
   The initial phase of an epidemic is often characterized by an exponential increase in the number of infected individuals.  In this paper, we predict the exponential spreading rate of an epidemic on a complex network. We first find an expression of the reproduction number for a network, based on the degree distribution, the network assortativity, and the level of clustering. We then connect this reproduction number and the disease infectiousness to the spreading rate. Our result holds for a broad range of networks, apart from networks with very broad degree distribution, where no clear exponential regime is present. Our theory bridges the gap between classic epidemiology and the theory of complex networks, with broad implications for model inference and policy making.
\end{abstract}

\maketitle
\section{Introduction}
Epidemic modeling permits the rationalization of evidence and provides crucial information for policy making. Epidemiological models should incorporate the key features that determine epidemic spreading.  For example, social structures play a crucial role in epidemic spreading  \cite{buckee2021thinking} and in the effectiveness of intervention strategies \cite{patwardhan2023epidemic,granell2018epidemic}. In these social structures, the number of contacts per individual is highly variable. For instance,  individuals with most contacts contributed disproportionately to the spreading of COVID-19 \cite{kumar2020significance,sneppen2020impact}. 

Social structures can be conveniently modeled by means of complex networks \cite{newman2018networks}, in which nodes represent individuals and links denote pairs of individuals that are in contact. A success of network theory has been to determine the epidemic threshold on networks \cite{pastor2001epidemic,newman2002spread,pastor2015epidemic}, i.e., the transition point from a localized outbreak to a widespread epidemic. This theory has shown that scale-free networks are extremely vulnerable to epidemics \cite{pastor2001epidemic,barthelemy2005dynamical,pastor2015epidemic,posfai2016network}.  Clustering, i.e., the tendency of nodes to form tightly connected communities, also affects epidemic spreading. Many real-world networks possess both heavy-tailed degree distribution and high levels of clustering \cite{ravasz2003hierarchical}.

Another key feature of an epidemic is its infectiousness, i.e., the rate at which an infected individual infects its contacts. The infectiousness incorporates multiple factors, such as viral load, social behavior, and possibly environmental factors \cite{grassly2008mathematical}. Because of changes in these factors, the infectiousness of an individual changes over time in a way that is particular to each disease. This time dependence profoundly impacts how a disease spreads.

Social structures and individual infectiousness are crucial in determining the initial stage of an epidemic, in which most individuals are susceptible and the number of infected individuals usually grows exponentially. This stage is of central importance for novel epidemics, or during the emergence of novel variants, as it often requires rapid policy making. At later stages, other factors become relevant, such as individual recovery time, their behavioral response \cite{weitz2020awareness}, and the possible occurrence of reinfections.

Despite the vast body of theory on epidemics in networks \cite{pastor2015epidemic}, general results on the speed of epidemic spreading are surprisingly sparse. The rate of epidemic spreading can be calculated for unclustered, uncorrelated networks, and with constant infectiousness  \cite{barthelemy2004velocity}. A recent theory estimates the average time it takes for an epidemic to reach an individual at a certain distance from patient zero \cite{moore2020predicting}. The spread of an epidemic on a network can be exactly solved \cite{merbis2022logistic}, but this calculation is practically feasible only for very small networks. Other approaches avoid a detailed description of social structures, and effectively model the population heterogeneity by assigning different risk propensities to individuals  \cite{rose2021heterogeneity,berestycki2023epidemic}. 

In this article, we theoretically predict the exponential rate of epidemic spreading on a complex network. Our theory quantifies the impact of the main properties of the network on the reproduction number of the epidemic and then links this reproduction number with the exponential spreading rate. We shall see that our approach leads to precise predictions for a broad range of networks. 

The manuscript is organized as follows. Section~\ref{sec:wellmixed} reviews classic concepts in epidemiology, in particular the definition of the reproduction number for well-mixed populations and its relation with the exponential spreading rate. Section~\ref{sec:model} introduces our model of epidemic spreading on networks. Section~\ref{sec:repr} discusses definitions of the reproduction number for networks, and shows how the reproduction number can be estimated. Section~\ref{sec:ELnetworks} contains our main result: the network Euler-Lotka equation, linking the reproduction number with the exponential spreading rate. Section~\ref{sec:weighted} extends our results to weighted networks. Section~\ref{sec:discussion} contains a discussion and future perspectives .

\section{Epidemic spreading in well-mixed populations}\label{sec:wellmixed}

We begin by summarizing how the exponential spreading rate of an epidemic is determined in well-mixed populations \cite{grassly2008mathematical}. 
We call $I(t)$ the average cumulative number of infected individuals at time $t$. The average number of newly infected individuals $\dot{I}(t)$ is governed by the renewal equation
\begin{equation}\label{eq:renew}
\dot{I}(t)=R_0\int_0^t d\tau\,  \omega(\tau) \dot{I}(t-\tau)\, ,
\end{equation}
where the basic reproduction number $R_0$ represents the average number of secondary infections by each infected individual. The function $\omega(\tau)$ represents the probability density that a given secondary infection occurs after a time $\tau$ from the primary one. Equation~\eqref{eq:renew} is linear because of the assumption that the epidemic is in the exponential stage and thus far from saturation.  We now substitute the ansatz $I(t)\sim e^{\Lambda t}$ into Eq.~\eqref{eq:renew} and take the limit of large $t$,  obtaining the Euler-Lotka equation
\begin{equation}\label{eq:EL}
\frac{1}{R_0}=\langle e^{-\Lambda \tau}\rangle\, ,
\end{equation}
where the average $\langle \dots \rangle$ is over the distribution $\omega(\tau)$.  Equation~\eqref{eq:EL} is a cornerstone in mathematical epidemiology, as it links the three key quantities governing the epidemic spreading: the distribution of infection times $\omega(\tau)$; the basic reproduction number $R_0$; and the exponential spreading rate $\Lambda$. 

\section{Model}\label{sec:model}

\begin{figure}[!ht]
    \centering
    \includegraphics[width=8cm]{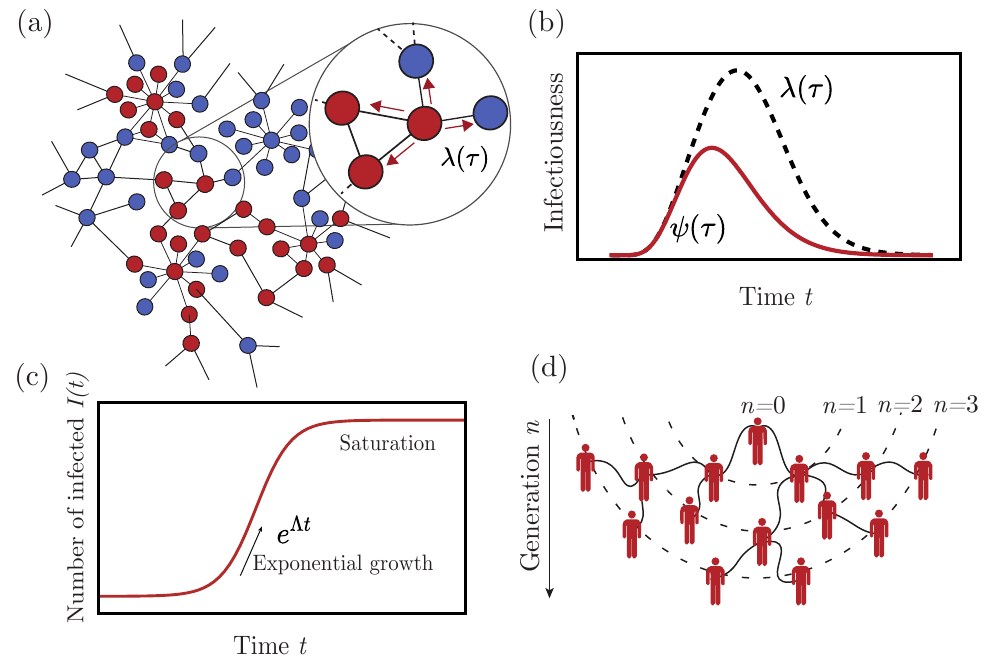}
    \caption{(a) Epidemic spreading on a network. Individuals (nodes) can be infected (red) or healthy (blue). Infected individuals spread the infection at a rate $\lambda(\tau)$ to each of their neighbors, where $\tau$ is the time since they were infected. If neighbors have already been infected, transmission does not result in an infection. (b) Time-dependent spreading rate $\lambda(\tau)$  and resulting distribution $\psi(\tau)$ of infection times. (c) An epidemic trajectory. The exponential phase ends in a saturation phase, where a sizable fraction of the  individuals are infected. (d) Epidemic tree. The epidemic initiates with one infected individual (patient zero). Subsequent infections are represented by continuous lines. Dashed lines represent  generations. }
    \label{fig:1}
\end{figure}

Our aim is to extend the concepts of Sec.~\ref{sec:wellmixed} to networks. Specifically, we consider a structured population composed of $N$ individuals, who are represented as nodes on a network [Fig.~\ref{fig:1}(a)]. Two individuals are connected by a link if they are in contact, meaning that they can potentially infect one another. An infected individual spreads the disease to each of its neighbors with rate $\lambda(\tau)$, where $\tau$ is the time since the individual was infected. The probability that the individual infects a given neighbor at time $\tau$, given that the neighbor has not been infected before by someone else, is then equal to $\psi(\tau)=\lambda(\tau)\exp[-\int_0^\tau \lambda(\tau')d\tau']$. We call $\psi(\tau)$ the generation time distribution [Fig.~\ref{fig:1}(b)]. The integral $p_\psi=\int_0^\tau \psi(\tau') d\tau'$ represents the total probability of infecting a susceptible neighbor. Aside from this normalization, the generation time distribution plays the role of $\omega(\tau)$ for networks. 

We assume that, over the timescale of interest, each individual can be infected at most once and never formally recovers, although their infectiousness may wane over time. Our model can thus be seen as a non-Markovian susceptible-infected (SI) model. We however remark that, in this non-Markovian framework, the distinction between SI and a susceptible-infected-recover (SIR) is only formal, as individuals whose infectivity $\lambda(\tau)$ has decayed to zero can be practically considered as recovered.

The time dependence of $\lambda(\tau)$ is important to model real-world epidemics. For example, the infectiousness of COVID-19 markedly increases for three to six days and then decays for about 11 days \cite{hakki2022onset}, and is therefore poorly described by a Markovian model with constant $\lambda(\tau)$.

In the model, the number of infected individuals grows in time until it saturates [Fig.~\ref{fig:1}(c)]. An alternative to the temporal perspective is to represent the epidemic in generations [Fig.~\ref{fig:1}(d)].  We call $Z_n$, $n=0, 1, \dots$ the average number of infected individuals in the $n$th generation. The initial generation, $n=0$, is constituted by the patient zero, $Z_0=1$. In this representation, the infected individuals form a tree [Fig.~\ref{fig:1}(d)].

\section{Reproduction number on a network}\label{sec:repr}

\subsection{Case $p_\psi=1$}

In this section, we discuss the definition of the reproduction number for a network. We first focus, for simplicity, on the case $p_\psi=1$, where individuals infect all of their contacts and the epidemic eventually spreads through the network, if the network is connected. 

The definition of the basic reproduction number $R_0$ for networks is slightly different from that in well-mixed populations. In well-mixed populations, $R_0$ is defined as the average number of secondary infections produced by a single infected individual in a fully susceptible population \cite{grassly2008mathematical}. Extending this definition to networks requires a specification of how the infecting node is selected. A node chosen uniformly at random will on average produce $\bar{k} = \sum_k kP(k)$ secondary infections, where $P(k)$ is the probability that the selected node has degree $k$. However, on a network infecting nodes are not chosen uniformly at random, but are rather infected by one of their neighbors. Consequently, susceptible nodes with higher degree are more likely to be infected. We therefore define $R_0$ as the average number of links of an infected individual in the early stage of the epidemic, excluding the link from which the considered individual received the infection. An equivalent definition is that  $R_0$ is the average degree (minus one) of the neighbor of a random node \cite{newman2002spread}. 

We also define the effective reproduction number $R$ as the average number of infections caused by an infected individual in the early stage of epidemics. In contrast to $R_0$, $R$ does not include contacts that are infected by other individuals. This distinction is necessary for networks, at variance with well-mixed populations. The reason is that, in well-mixed populations, most contacts of an individual are susceptible if the fraction of infected individuals is low. However, this is not necessarily the case in networks, as the neighborhood of an infected individual can be locally saturated even if the global fraction of infected individuals is low.

To estimate $R$ for a network, we characterize individuals by their degree $k$. We introduce the reproduction matrix
\begin{equation}\label{eq:matrixdiag}
M_{kk'} = (k'-1-m_{kk'})P(k|k') \, ,  
\end{equation}
whose elements are the average numbers of susceptible individuals of degree $k$ connected to a node of degree $k'$. In Eq.~\eqref{eq:matrixdiag}, $P(k|k')$ is the probability that a neighbor of a node of degree $k'$ has degree $k$ and $m_{kk'}$ is the average number of triangles a $kk'$-link is part of. The term proportional to $m_{kk'}$ discounts for contacts of a given individual that were already infected by someone else, due to the presence of triangles. In Eq.~\eqref{eq:matrixdiag}, we neglect higher-order loops such as squares. The reproduction matrix  has been previously used to study percolation on networks \cite{serrano2006clusteringii}. 

We express the average number $Z_n$ of infected individuals at generation $n\ge1$ in terms of $M_{kk'}$ by summing over the degrees of individuals at each generation: 
\begin{equation}
    Z_n=\sum_{k_1\dots k_n}M_{k_n k_{n-1}}M_{k_{n-1} k_{n-2}}\dots M_{k_2 k_{1}}k_1P(k_1)\, .
\end{equation} 
At the first generation, we obtain $Z_1=\bar{k}$. For large $n$, we have $Z_n\sim R^n$, where we identify the reproduction number $R$ as the leading eigenvalue of $M_{kk'}$. The associated right eigenvector $v_{k'}$ represents the degree distribution of infected individuals, see Appendix \ref{sec:pertub_derivation}.

In the simple case of an unclustered network $(m_{kk'}=0$ for all links) in which the degrees of connected nodes are uncorrelated, we have
\begin{equation}\label{eq:R0}
R=R_0=\frac{\av{k^2} -\av{k}}{\av{k}} \, ,
\end{equation}
where the bar represents an average over the degree distribution. Equation~\eqref{eq:R0} is a classic result \cite{pastor2001epidemic,callaway2000network,newman2002spread}. 

For clustered and correlated networks, $R_0$ is still given by the right-hand side of Eq.~\eqref{eq:R0}, while the value of $R$ may in general differ. To estimate this value, we quantify degree correlations between connected nodes by the assortativity coefficient $r$, see Appendix \ref{sec:pertub_derivation}. This coefficient measures the tendency for nodes in a network to connect preferentially to others with similar degrees: When $r>0$ high-degree nodes tend to link to other high-degree nodes, while for $r<0$ high-degree nodes are more likely to connect with low-degree nodes. When $r=0$, degrees of connected nodes are independent, and thus the network is uncorrelated, following $P(k|k')=k/\bar{k} P(k)$ \cite{pastor2015epidemic}.  Our approach is to perturbatively expand $P(k|k')$ around the uncorrelated case at the first order in $r$, see Appendix~\ref{sec:pertub_derivation}. We find that the effective reproduction number is approximated by
\begin{equation}
R\approx R_\pert=(1-r)(R_0-m_{1})+
r\left(\frac{\frac{\av{ k(k-1)^2}}{\av{k}}-2m_{2} +m_1^2}{R_0-m_1}\right) \,,
\label{eq:R_tilde}
\end{equation}
where $c_k$ is the probability that two neighbors of a node of degree $k$ are themselves connected, $m_{1}=\overline{k(k-1) c_k}/\overline{k}$ is the average number of triangles a link is part of, and $m_{2}=\overline{k(k-1)^2 c_k}/\overline{ k}$, see Appendix \ref{sec:pertub_derivation}. 

\begin{table}[htb]
    \caption{Reproduction numbers. Here, $R$ is the numerically computed leading eigenvalue of Eq.~\eqref{eq:matrixdiag}. The expressions of $R_0$ and $R_\pert$ are given by Eq.~\eqref{eq:R0} and Eq.~\eqref{eq:R_tilde}, respectively. To assess the impact of assortativity and clustering, we also compute $R_\pert$  for $r=0$ and for $\bar{c}=\sum_k c_k P(k)=0$.  BA$_m$: Barab\'asi-Albert with parameter $m$. $\text{ER}_d$: Erd\H{o}s-R\'enyi with mean degree $d$. WS$_{k,p}$: Watts-Strogatz starting from a 1D lattice with $k$ nearest neighbors and rewiring probability $p$. Additional examples are given in \cite{suppl}.\label{tab:R_comparison}}
\begin{ruledtabular}
    \begin{tabular}{cccccccc}
    \text{Network} & $R_0$ & $R$ & $R_\pert$ & $R_\pert^{(r=0)}$ & $R_\pert^{(\av{c}=0)}$ & $r$ &  $\av{c}$ \\ \midrule
    $\text{BA}_1$ & 13.9 & 8.5 & 6.8 & 13.9 & 6.6 & -0.011 & 0 \\
    $\text{BA}_2$ & 20.4 & 15.5 & 14.4 & 20.3 & 14.2 & -0.008 & 0.0001 \\
    $\text{BA}_{12}$ & 85.4 & 80.8 & 79.5 & 83.6 & 80.4 & -0.004 & 0.0003 \\
    $\text{ER}_{6}$ & 6.0 & 6.0 & 6.0 & 6.0 & 6.0 & 0.001 & 0.000 \\
    $\text{WS}_{8,2}$ & 7.2 & 4.9 & 4.9 & 4.9 & 7.0 & -0.022 & 0.334 \\
    $\text{WS}_{6,2}$ & 5.2 & 3.6 & 3.6 & 3.6 & 5.0 & -0.029 & 0.315 
    \end{tabular}
\end{ruledtabular}
\end{table}

We compare $R_\pert$ given by Eq.~\eqref{eq:R_tilde} with the maximum eigenvalue $R$ of the matrix $M_{kk'}$ for different network models, see Table~\ref{tab:R_comparison}.  The relative difference between $R$ and $R_\pert$ is within $2\%$ for all the models we considered, except for Barab\'asi-Albert with parameter $m=1$ ($20.6\%$) and $m=2$ ($7.1\%$), where $m$ represents the number of nodes a new node attaches to. This is likely due to the perturbative term being large in these two cases. Approximating $R$ with $R_0$ as given by Eq.~\eqref{eq:R0} leads to substantially larger errors, supporting that clustering and assortativity play an important role. Here and in the following, when computing $R_{\pert}$ using Eq.~\eqref{eq:R_tilde} and $R_0$ using Eq.~\eqref{eq:R0} for a given network, we always interpret the averages as empirical averages. With this choice, moments of the degree distribution remain finite even in very heterogeneous networks.  

\begin{figure}[!h]
    \centering
    \begin{subfigure}[b]{0.23\textwidth}
        \centering
        \includegraphics[width=\linewidth]{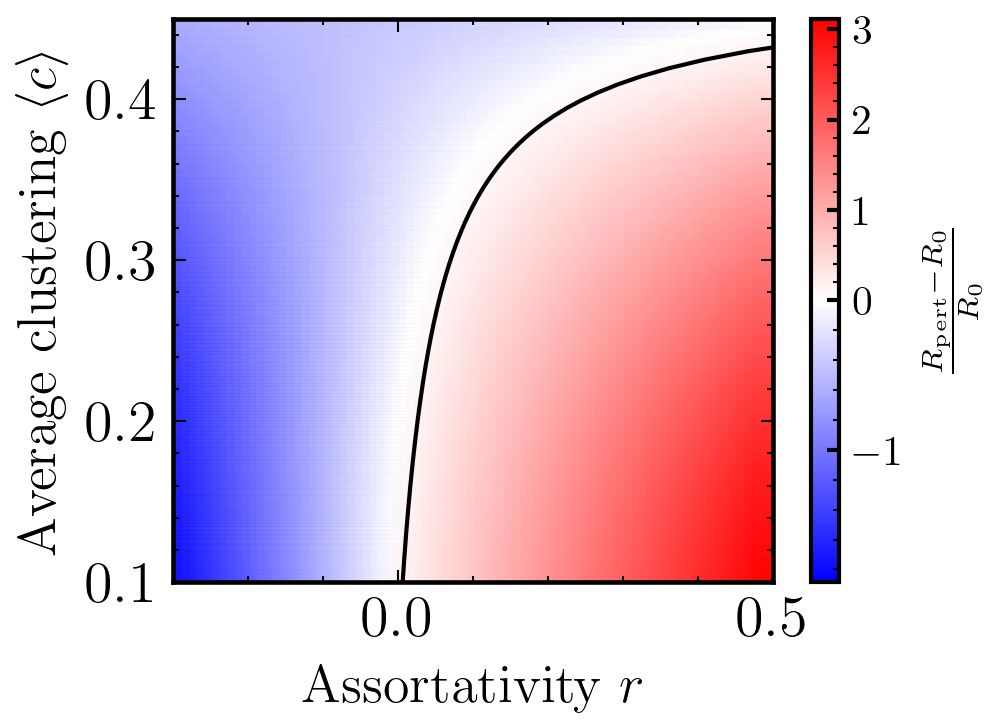}
        \caption{}
    \end{subfigure}
    \centering
    \begin{subfigure}[b]{0.23\textwidth}
        \centering
        \includegraphics[width=\linewidth]{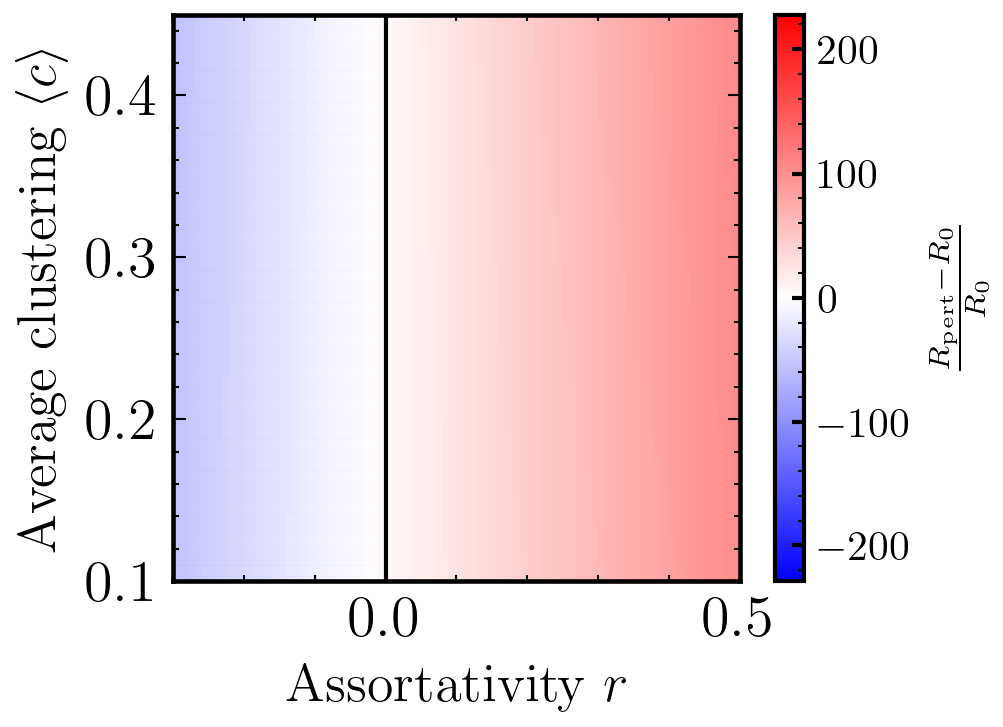}
        \caption{}
    \end{subfigure}
    \caption{Phase diagram representing the interplay between assortativity and clustering on the reproduction number for different degree distributions. We numerically compare the formulas for $R_0$,  Eq.~\eqref{eq:R0}, and $R_{\rm pert}$, Eq.~\eqref{eq:R_tilde}, with (a) a Poisson distribution with mean $5$  and (b) with a power-law with an exponential cutoff of $k_c=300$ ($P(k)\propto k^{-3}e^{-k/k_c}$).  We assume that $c_k=m_0 (k-1)^{-\alpha}$ and we compute $R_{\rm pert}$ using Eq.~\eqref{eq:R_tilde}, for various values of $m_0$ and $\alpha$. The black curve represents the set of points characterized by  $R_{\pert}=R_0$. The blue region represents the regime where $R_{\rm pert}$ is lower, whereas the red region represents the regime where $R_{\rm pert}$ increases the reproduction number compared to $R_0$.}
    \label{fig:phasediagram}
\end{figure}

In practical cases, Eq.~\eqref{eq:R_tilde} predicts that $R_\pert$ increases with the assortativity $r$. Accordingly, assortativity lowers the epidemic threshold, as previously known for the unclustered case \cite{goltsev2012localization}. On the contrary, clustering substantially decreases $R_\pert$. Although assortativity and clustering influence the reproduction number in opposing ways, clustering typically exerts a stronger impact. This explains why $R_0$ consistently overestimates the effective reproduction number $R$ across all synthetic networks considered in our analysis. An exception occurs in networks with heavy-tailed degree distributions, where the influence of assortativity is significantly amplified, as $r$ approximately scales with $\av{k^3}$ in Eq.~\eqref{eq:R_tilde}, and may outweigh clustering effects (see Fig.~\ref{fig:phasediagram} for an example).

\subsection{Case $p_\psi<1$}
\label{sec:p_psi}

We now move to the case in which infected nodes do not always transmit the infection to all their contacts, meaning that $p_\psi=\int_0^\tau \psi(\tau') d\tau'<1$. 
In this case, we define a reproduction number $R^{\psi}$ as the average number of secondary infections per infected individuals. While $R$ depends on the topology of the network only, $R^\psi$
depends on $p_\psi$ as well. We have $R^{\psi}=R$ in
the limiting case $p_\psi=1$.

In our model, the infection spreads through every link at most once. A case in which the infection fails to spread via a specific link is equivalent to the case in which the link did not exist in the first place. This means that the scenario $p_\psi<1$ can be studied by removing links with uniform probability $(1-p_\psi)$. This procedure alters the degree distribution of the network, which becomes
\begin{equation}
    P^{\psi}(k) = \sum_q  \binom{q}{k}{p_\psi}^k \left(1-p_\psi \right)^{q-k}P(q).
    \label{eq:bond_perc}
\end{equation}
When the network is unassortative and unclustered, the reproduction number scales linearly with $p_\psi$:
\begin{align}\label{eq:rpsi_uncorr}
    \sum_k (k^2-k)P^{\psi}(k)&={p_\psi}^2\left(\av{k^2}-\av{k}\right)\\
    \sum_k kP^{\psi}(k)&={p_\psi}\av{k}\\
    \implies R^{\psi} = p_\psi R,
\end{align}
where $R=\av{k^2}/\av{k}-1$, and the overbar still denotes averages over $P(k)$. Thus, the epidemic spreading condition ($R^\psi>1$) implies
\begin{equation}
    p_\psi>\frac{\av{k}}{\av{k^2}-\av{k}}.
\end{equation}
This relation recovers the classic result for the epidemic threshold on uncorrelated networks \cite{cohen2000resilience,newman2002spread}.

This linear relation given in Eq.~\eqref{eq:rpsi_uncorr} does not hold, in general, for clustered or assortative networks. Triangles are vulnerable to link removal since all three links need to remain for a triangle to be preserved. If a given link belongs on average to $m_1$ triangles, after link removal it would belong on average to only $m_1{p_\psi}^2$ triangles. Therefore, the reproduction number, in the absence of degree correlations, is expressed by
\begin{equation}\label{eq:Rnew}
    R^{\psi} = p_\psi R_0 - m_1{p_\psi}^2.
\end{equation}
We now consider the effect of assortativity. Under edge removal with uniform probability $1-p_\psi$, the assortativity $r$ is transformed into 
\begin{equation}\label{eq:rnew}
    r^{\psi}=r\left(1+\frac{p_\psi}{1-p_\psi}\frac{R_0}{\sigma_r^2}\right)^{-1},
\end{equation}
where $\sigma_r^2=\av{k^3}/\av{k}-\left(\av{k^2}/\av{k}\right)^2$. Equation~\eqref{eq:rnew} is derived in Ref.~\cite{noh2007percolation}. Similarly to the uncorrelated case, we obtain
\begin{equation}
\frac{\sum_k P^{\psi}(k)k(k-1)^2}{\sum_{k'} k'P^{\psi}(k')}= p_\psi^2\frac{\av{k(k-1)^2}}{\av{k}}+R_0 p_\psi(1-p_\psi)
\end{equation}
and
\begin{equation}
m_2^{\psi}\to p_\psi^3 m_2+p_\psi^2(1-p_\psi)m_1 .
\end{equation}
Therefore, we have
\begin{align}\label{eq:rpert_perco}
    R_\pert^{\psi} &= p_\psi \left(R_0 - m_1{p_\psi})\right)(1-r^{\psi})+r^{\psi}(1-p_\psi)+\nonumber\\
    &\frac{r^{\psi}}{R_0 - m_1{p_\psi}}\left(p_\psi\frac{\av{k(k-1)^2}}{\av{k}}-2p_\psi^2 m_2\right.\nonumber \\
    &\quad-p_\psi m_1(1-p_\psi+m_1p_\psi^2)\biggr)+\mathcal{O}(r^2)\, ,
\end{align}
where $r^{\psi}$ is given by Eq.~\eqref{eq:rnew}.

\section{Network Euler-Lotka equation}\label{sec:ELnetworks}

We now use our estimates of $R$ to predict how epidemics spread in time on a network. The average number $\dot{I}_k(t)$ of newly infected individuals of degree $k$ at time $t$ is governed by the renewal equation
\begin{equation}
    \dot{I}_k(t)=\sum_{k'}\int_0^t  \dot{I}_{k'}(t-\tau)M_{kk'}\psi(\tau)\mathrm{d}\tau\, .
\label{eq:renewal}
\end{equation} 
Equation~\eqref{eq:renewal} descends from the definitions of $M_{kk'}$ and $\psi(\tau)$, and extends Eq.~\eqref{eq:renew} to complex networks. Assuming exponential growth, $\dot{I}_{k'}(u)\propto v_{k'}e^{\Lambda t}$, we obtain for large time
\begin{equation}\label{eq:netEL}
\frac{1}{R}=\left\langle e^{-\Lambda \tau}\right\rangle \, ,
\end{equation}
where $\langle\dots\rangle$ denotes an average over $\psi(\tau)$.  We call Eq.~\eqref{eq:netEL} the network Euler-Lotka equation. See Appendix \ref{sec:LDP} for an alternative derivation using large deviation theory.

We shall use Eq.~\eqref{eq:netEL} as an implicit relation for the unknown $\Lambda$, and call $\Lambda_{\mathrm{EL}}$ the solution of the equation. In some cases, the equation can be explicitly solved. For example, in the Markovian limit (constant infectiousness $\lambda(t)=\lambda$ and $\psi(t)=\lambda e^{-\lambda t}$), Eq.~\eqref{eq:netEL} predicts that $\Lambda=\lambda(R-1)$. In general, we can only explicitly express $\Lambda$ from Eq.~\eqref{eq:netEL} in an approximate way. In particular, using the first two moments of $\psi(\tau)$, we obtain 
\begin{equation}
    \Lambda \approx \frac{\ln R}{\langle \tau \rangle}+\left(\ln R\right)^2\frac{\sigma^2}{2\langle\tau\rangle^3}\, ,\label{eq:lambda_approx}
\end{equation}
where $\sigma^2=\langle \tau^2\rangle - \langle \tau\rangle^2$, see Appendix \ref{sec:LDP}. An alternative is to combine Eq.~\eqref{eq:netEL} with the Jensen inequality, that states that $\langle  e^{-\Lambda \tau}\rangle \ge e^{-\Lambda \langle \tau\rangle}$. This leads to the lower bound  
\begin{equation}
\Lambda \ge (\ln R)/\langle \tau \rangle \, .
\end{equation}
This bound is saturated when $\psi(\tau)$ is a Dirac delta function.

\subsection{Synthetic networks} 
\begin{figure}[htb]
    \centering
\includegraphics[width=8cm]{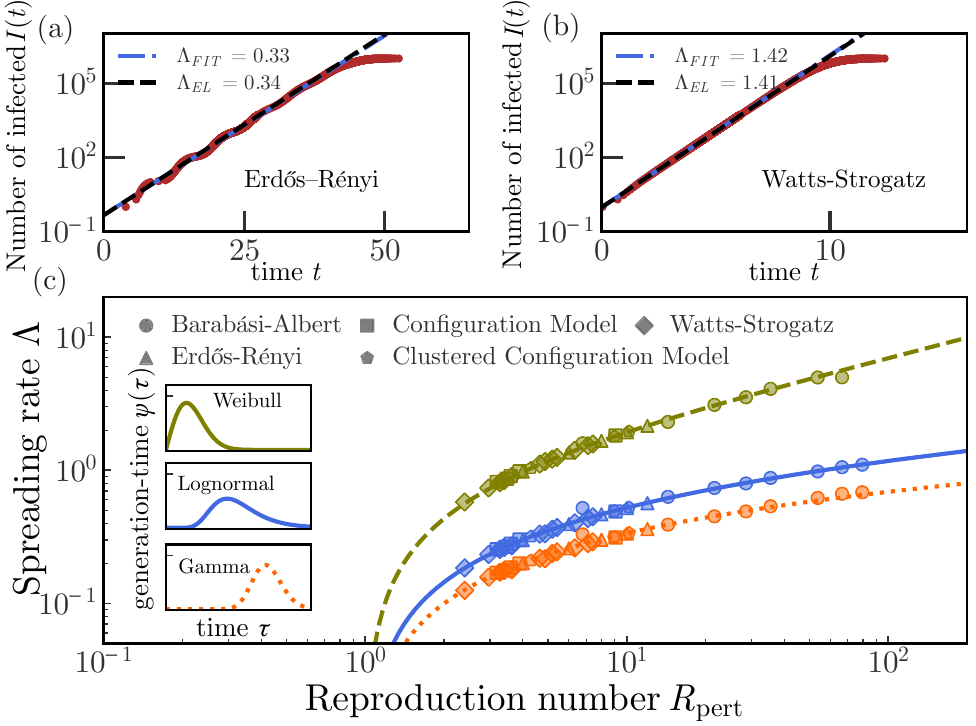}
         \caption{Network Euler-Lotka equation predicts the epidemic spreading rate on synthetic networks. (a) and (b): Average trajectory $I(t)$ over 500 realizations (red), $\Lambda_\text{FIT}$ (blue dot-dashed) and $\Lambda_\text{EL}$ (black dashed; based on $R_\text{pert}$, Eq.~\eqref{eq:R_tilde}) for Erd\H{o}s-R\'enyi ($N=10^6$, $\av{k}=4$) network with Gamma-distributed infection times and Watts-Strogatz network with Weibull-distributed infection times. (c) Solution of Eq.~\eqref{eq:netEL} compared with simulations for various networks of sizes up to $10^6$, see Appendix \ref{sec:simulations} and \cite{cure2024fast}. 
     }
    \label{fig:2}
\end{figure}
The network Euler-Lotka equation predicts the rate of epidemic spreading on synthetic complex networks with remarkable accuracy, Figs.~\ref{fig:2}(a) and ~\ref{fig:2}(b).  Our battery of tests includes a range of models characterized by different degree distribution, assortativity, and clustering. For each model, we use three different generation time distributions $\psi(\tau)$, see Fig.~\ref{fig:2}(c). All three distributions are characterized by $p_\psi=1$; we shall consider the case $p<\psi<1$ in Sec.~\ref{subsec:el_psi}. The choice of $\psi(\tau)$ qualitatively affects the epidemic spreading. In particular, for peaked distributions, the exponential growth appears modulated by oscillations [Fig.~\ref{fig:2}(a)], since early generations of the epidemic are nearly synchronized. To extrapolate the leading exponential behavior from these curves, we fit them into a generalized logistic function, 
\begin{equation}
   I(t) = \frac{N}{\left(1+\left((N/I_0)^\nu-1\right)e^{-\Lambda_0\nu t}\right)^{1/\nu}}\, ,
   \label{eq:logi}
 \end{equation}
which appears to well capture finite-size effects. In Eq.~\eqref{eq:logi}, $N$ the size of the network and $I_0$, $\Lambda_0$, and $\nu$ are free parameters. We fit the logarithm of $I(t)$ versus $t$ and discard the initial times at which $I(t)<\av{k}$. The reason is that the first $\av{k}$ infections should belong to the first generation, in which the average number of secondary infections per individual is $Z_1=\av{k}$ rather than $R$. Although each simulation starts with a single infected individual, we consider $I_0$ as a free parameter to better capture the behavior of the epidemic at the early stages. Expanding Eq.~\eqref{eq:logi} for small $t$, we obtain 
\begin{equation}
\log I(t)\approx\log I_0 + t\Lambda_0 [1-(I_0/N)^{\nu }] \, .
\end{equation}
This leads us to define the finite-size growth rate 
\begin{equation}
\Lambda_{\rm FIT}=\Lambda_0(1-\eta)\, ,
\end{equation}
and the finite-size parameter 
\begin{equation}
\eta=\left(I_0/N\right)^{\nu }\in [0,1]\, ,
\end{equation}
which quantifies the impact of finite-size effects on the exponential spreading rate.

When computing $\Lambda_{\mathrm{EL}}$, we set $R=R_\text{pert}$ as defined in Eq.~\eqref{eq:R_tilde} as the reproduction number. As expected, the predictions are considerably worse if we instead use $R_0$, see Tables III-VIII in Supplementary Material \cite{suppl} and Appendix \ref{sec:R_vs_R0}. The difference between $\Lambda_{\mathrm{FIT}}$ and $\Lambda_{\mathrm{EL}}$ is within $6\%$  for all networks, see Fig.~\ref{fig:2}(c). The only exception is the Barab\'asi-Albert model with parameter $m=1$ (average error $15\%$). One explanation for this discrepancy is that the exponential regime for this model is not well defined for $N\rightarrow \infty$, since $\bar{k^2}$ diverges in this limit. Another explanation is that, for $m=1$, $R_\pert$ does not accurately match the leading eigenvalue of the reproduction matrix (see Table~\ref{tab:R_comparison}).

\begin{figure}[htb]
	\centering
    \includegraphics[width=8cm]{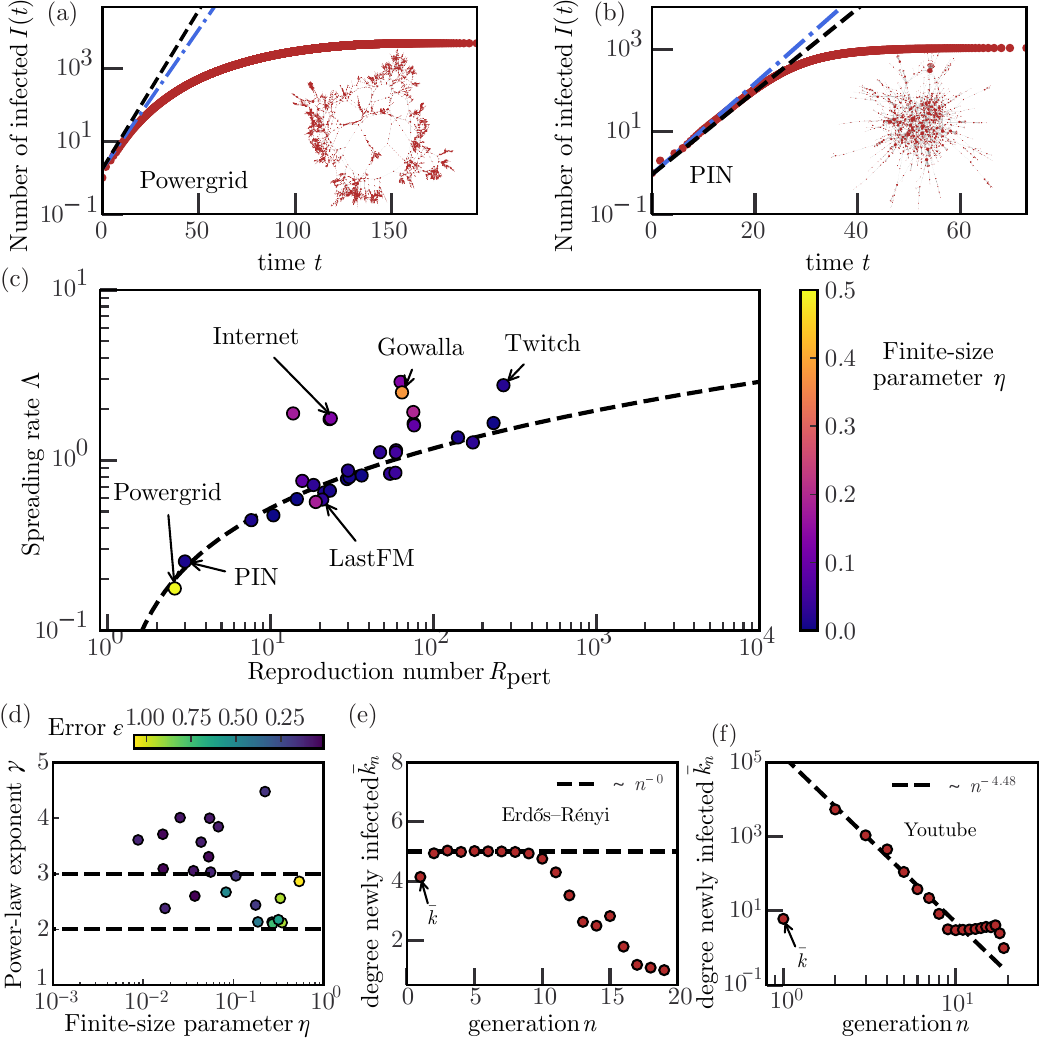}
	\caption{Network Euler-Lotka equation predicts the epidemic spreading rate on real-world networks. We obtained $I(t)$ by averaging $500$ simulations using log-normal infection times for each network from Refs.~\cite{clauset2016colorado,snapnets,konect} with at least $1000$ nodes \cite{suppl}. In all cases, we have set $p_\psi=1$. (a) and (b): Average trajectory $I(t)$ (red), $\Lambda_\text{FIT}$ (blue dot-dashed), and $\Lambda_\text{EL}$ (black dashed) for power grid and PIN networks. (c) Exponential spreading rate $\Lambda_\text{EL}$ (dashed line) versus $\Lambda_\text{FIT}$ (markers). (d) Degree distribution tail exponent ($\gamma)$ and finite-size parameter ($\eta$) affect the estimation error $\varepsilon= 2| \Lambda_\text{FIT}-\Lambda_{\rm EL}|/(\Lambda_\text{FIT}+\Lambda_{\rm EL})$. Each dot represents a network, dashed lines mark the scale-free region $2<\gamma\leq 3$. (e) and (f) Average degree $\bar{k}_n$ of infected individuals at $n$-th generation for (a) Erd\H{o}s-R\'enyi ($N=10^6$, $\av{k}=4$) and (b) YouTube (scale-free, $\gamma \approx 2.2$). Dashed lines represent power-law fits for $n\in[1,6]$.}
	\label{fig:3}
\end{figure}

\subsection{Real-world networks}
We apply the network Euler-Lotka equation to a large set of social, biological, technological, and transportation networks. These real-world networks often exhibit more complex structures than synthetic networks, with tightly linked communities, strong degree correlation, and broad tails in the degree distribution. Although they constitute a challenging test, the network Euler-Lotka equation holds well, see Figs.~\ref{fig:3}(a) ~\ref{fig:3}(b) for examples. More generally, we find that $\Lambda_\text{EL}$ well matches $\Lambda_\text{FIT}$ whenever the finite-size parameter $\eta$ is small, see Fig.~\ref{fig:3}(c). 

The fact that $\eta$ is small implies the presence of a clear exponential regime. To show that, we express the variation of the instantaneous exponential growth rate over a given timescale $\delta t$ as $d^2 \ln I(t)/dt^2|_{t\rightarrow 0+}\delta t$. Substituting Eq.~\eqref{eq:logi}, this implies that the relative error on the exponent can be estimated as 
\begin{equation}
-\frac{\delta t}{\Lambda}\left.\frac{d^2 \ln I(t)}{dt^2}\right|_{t\rightarrow 0+}=\nu\eta\Lambda_0\delta t \, ,
\end{equation}
which is proportional to $\eta$. For example, in the case of the power grid network, we have $\eta=0.5$, $\nu=0.09$, and $\Lambda_0=0.35$. In this case, the relative error becomes of order one over a timescale $\delta t\sim (\nu\eta\Lambda_0)^{-1}\approx 60$ days, which is before the epidemic has reached saturation, see Fig.~\ref{fig:3}(a). Thus the parameter $\eta$ also serves as a measure of deviations from an exponential behavior.

Although some of these networks are highly clustered, using $R_\pert$ as expressed by Eq.~\eqref{eq:R_tilde} does not perform worse than $R$, see Tables in \cite{suppl}. In Fig.~\ref{fig:3}c, we have omitted three road networks, for which the fit of Eq.~\eqref{eq:logi} fails to converge. However, a function of the form $\dot{I}(t)\sim t^{\beta}$ well fits the epidemic trajectories in these cases, see Appendix \ref{power-law-fit}. This is likely because these networks are embedded in a two-dimensional physical space, see \cite{moore2024network}.
For networks in which $\eta<0.5$, the predictions of Eq.~\eqref{eq:netEL} using $R_\text{pert}$ are more accurate than those using $R_0$. However, for larger values of $\eta$, $R_0$ performs slightly better on average. Even in these cases, $R_\pert$ provides a better approximation of $R$, see Appendix \ref{sec:R_vs_R0}. The advantage of $R_0$ for large $\eta$ might be due to a compensatory effect: using $R_0$ tends to overestimate $R$, while the network Euler-Lotka equation tends to underestimate $\Lambda$ compared to $\Lambda_{\mathrm{FIT}}$.

All of the networks where the prediction error is large are scale-free and characterized by a large value of $\eta$, see Fig.~\ref{fig:3}(d). Here, scale-free means that the tail of the degree distribution scales as $k^{-\gamma}$, with $\gamma\le3$. Specifically, a degree distribution is considered heavy-tailed or scale-free if it can be written in the form $P(k)=\ell(k)k^{-\gamma}$, where $\ell(k)$ is a slowly varying function satisfying $\lim_{k \rightarrow \infty} \ell(xk)/\ell(k)=1$, see \cite{voitalov2019scale,broido2019scale}. This generalizes beyond pure power-law distributions, accounting for real-world data often better fit by distributions like lognormal or those associated with models such as the \BA, which do not follow a pure power law\cite{posfai2016network}. We estimate the exponent $\gamma$ following the approach from Ref.~\cite{voitalov2019scale}. The method estimates an extreme value index $\xi = (\gamma - 1)^{-1}$ using kernel estimators \cite{groeneboom2003kernel, voitalov2018code}. If the kernel method fails, we resort to the method of moments. The specific values of $\gamma$ estimated for each real network using different methods are provided in the Supplementary Material \cite{suppl}. 

\begin{figure}[!ht]
    \centering
    \includegraphics[width=0.44\textwidth]{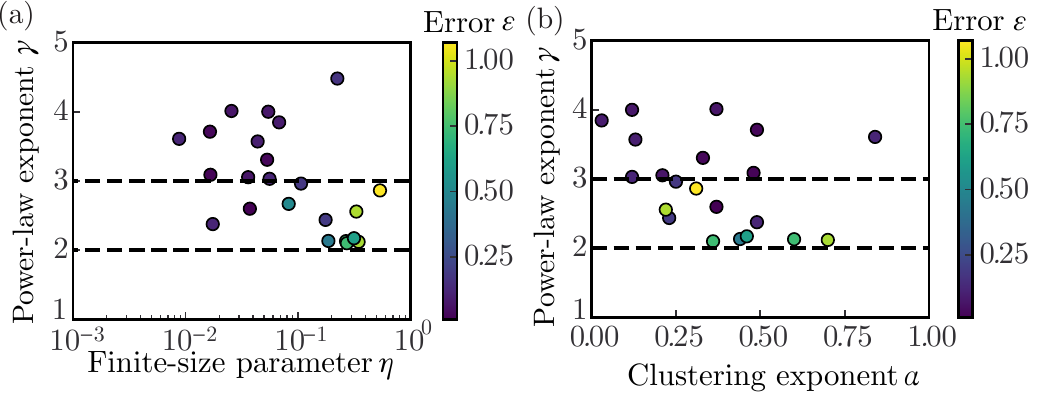}
    \caption{ Effect of degree distribution tail exponent ($\gamma)$, finite-size parameter ($\eta$) and clustering coefficient ($\alpha)$ on the estimation error $\varepsilon= 2| \Lambda_{\rm FIT}-\Lambda_{\rm EL}|/(\Lambda_{\rm FIT}+\Lambda_{\rm EL})$. Each dot represents a network, dashed lines mark the scale-free region $2<\gamma\leq 3$. (a) Effect of $\gamma$ and $\eta$ on $\epsilon$  [identical to Fig.~\ref{fig:3}(d)]. (b) Effect of $\gamma$ and $\alpha$ on $\epsilon$.}
    \label{fig:fig3_d2}
\end{figure}

The error for scale-free networks can be explained by the early infection of hubs, which depletes the tail of the degree distribution. In networks that are not scale-free, we find that the average degree of infected individuals is constant among generations as expected [Fig.~\ref{fig:3}(e)]. In contrast, we find  this average degree to decay as a power law in scale-free networks [Fig.~\ref{fig:3}(f)], leading to a nonsteady spreading. We also tested whether clustering affects the accuracy of the predicted spreading rate and found no significant effect, see Fig.~\ref{fig:fig3_d2}.

\subsection{Case $p_\psi<1$}\label{subsec:el_psi}

\begin{figure}[!h]
    \centering
    \includegraphics[width=0.33\textwidth]{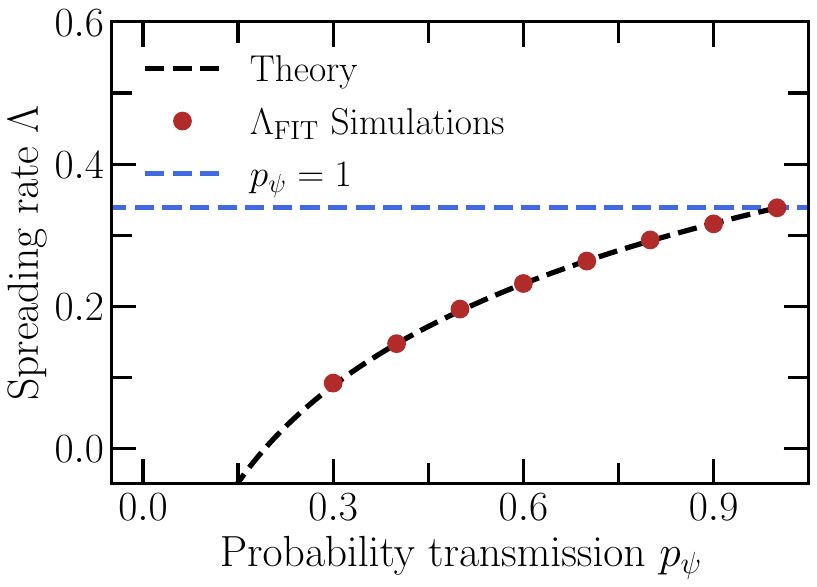}
    \caption{ Effect of non-normalized generation-time on the spreading rate. We generate a \WS network with parameter $k=6$ and rewiring probability $p=0.3$. We then simulate the average epidemic trajectory over $100$ trajectories using a Gamma generation-time distribution, imposing that the transmission via each link occurs with probability $p_\psi$. We repeat this for different values of $p_\psi$ and measure the empirical spreading rate (red markers). The black dashed line is the prediction of Eq.~\eqref{eq:Rnew}. The blue dashed line marks the spreading rate for $p_\psi=1$.}
    \label{fig:percolation_vs_growth_rate}
\end{figure}

We numerically test the case $p_\psi<1$ on a \WS network, which is clustered and has an almost vanishing correlation in the connected degrees ($r\approx 0)$. Our simulations are in excellent agreement with the predictions of Eq.~\eqref{eq:Rnew}, see Fig.~\ref{fig:percolation_vs_growth_rate}, provided that $p_\psi$ is large enough so that the decimated network is well above its percolation threshold and therefore guaranteed to have a giant component. 

\begin{figure}[!hb]
    \centering
    \begin{subfigure}[b]{0.23\textwidth}
        \centering
        \includegraphics[width=\textwidth]{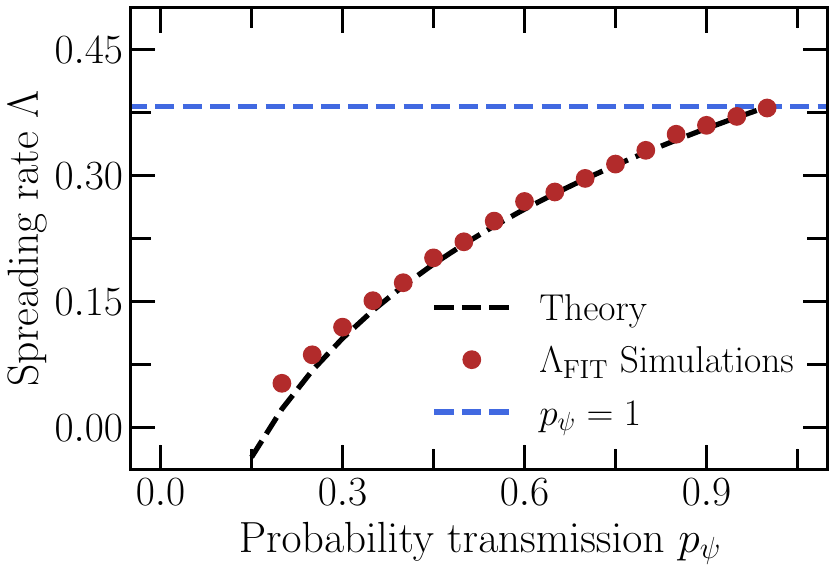}
        \caption{Lognormal network}
        \label{fig:percolation_lognorm}
    \end{subfigure}
    \begin{subfigure}[b]{0.23\textwidth}
        \centering
        \includegraphics[width=\textwidth]{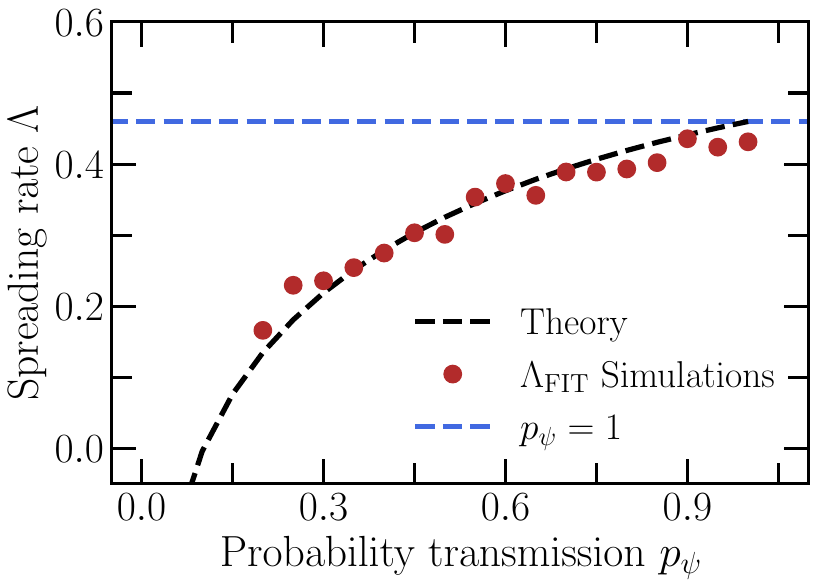}
        \caption{Amazon network}
        \label{fig:percolation_amazon}
    \end{subfigure}
    \caption{Impact of the transmission probability on the epidemic spreading rate. For each value of $p_\psi$ (red markers), we simulate the average epidemic trajectory over $100$ trajectories with infection times that are Gamma distributed. The chosen networks are :(a) a configuration model in which the degree is lognormally distributed, (b) the Amazon network. The black dashed line represent the solution of the Euler-Lotka equation with $R=R^{\psi}_\pert$ given by Eq.~\eqref{eq:rpert_perco}.}
\end{figure}

To test Eq.~\eqref{eq:rpert_perco}, we simulate epidemics with a Gamma generation-time distribution while varying $p_\psi$ on two different networks. First, we consider a synthetic network that is disassortative and exhibits a heavy-tail in its degree distribution, see Fig.~\ref{fig:percolation_lognorm}. Second, we consider the Amazon network, where nodes represent products and edges link commonly copurchased products \cite{snapnets,leskovec2007dynamics}. The Amazon network  presents assortativity, a heavy tail, and clustering. We find that our theory is in excellent agreement with the simulations provided that $p_\psi$ is not too small, as expected, see Fig.~\ref{fig:percolation_amazon}. 

\section{Weighted networks}\label{sec:weighted}

To realistically model epidemic spreading, it is often necessary to assume that interactions between individuals are not always of the same type. For instance, infected individuals might spread the disease more rapidly to contacts in closed environments, such as workplaces, compared to outdoor spaces. To account for this, we generalize the network Euler-Lotka equation to networks in which edges are assigned a type $i$ with corresponding weight $w_i$ and transmission time distribution 
\begin{equation}
\psi^{(i)}(\tau) =  w_i \lambda(\tau)\exp\left[-\int_0^\tau w_i \lambda(\tau')d\tau'\right]\, .
\end{equation}
In this case, the renewal equation becomes
\begin{equation}
    \label{eq:multitype_renewal}
    \dot I_k(t) = \sum_i \sum_{k'} \int_0^\infty M_{kk'}^{(i)} \dot I_{k'}(t - \tau) \psi^{(i)}(\tau) d\tau ,
\end{equation}
where we have defined 
\begin{equation}
  \label{eq:multitype_M_kk_i}
  M_{kk'}^{(i)} = Q(i|k, k') M_{kk'}
\end{equation}
where $Q(i|k,k')$ is the fraction of edges with weight $w_i$ among the edges connecting nodes of degree $k$ and $k'$. Since $\sum_i Q(i|k,k')=1$ by construction, we have that $M_{kk'} = \sum_i M_{kk'}^{(i)}$. Assuming exponential growth, $\dot{I}_{k'}(u)\propto v_{k'}e^{\Lambda t}$ for large times, we obtain the weighted Euler-Lotka equation,
\begin{equation}
    \label{eq:multityp_EL}
    v_k = \sum_{k'} \left(\sum_i \big\langle e^{-\Lambda \tau}\big\rangle_i M_{kk'}^{(i)} \right) v_{k'},
\end{equation}
where $\langle \cdots \rangle_i = \int_0^\infty \cdots \ \psi^{(i)} d\tau$ denotes the average over $\psi^{(i)}(\tau)$. Defining the matrix
\begin{equation} \label{eq:tildeM}
\tilde{M}_{kk'}(\Lambda)=\sum_i \big\langle e^{-\Lambda \tau}\big\rangle_i M_{kk'}^{(i)}\, ,
\end{equation}
the weighted Euler-Lotka equation imposes that the exponential growth rate is the value of $\Lambda$ such that the leading eigenvalue of $\tilde{M}_{kk'}(\Lambda)$ is equal to one. The derivation can be readily extended to a continuous set of weights by replacing the sums over $i$ with integrals; in this case, the function $Q$ becomes a probability density over the weights. 

\begin{figure}
    \centering
    \begin{subfigure}[b]{0.23\textwidth}
    \includegraphics[width=\linewidth]{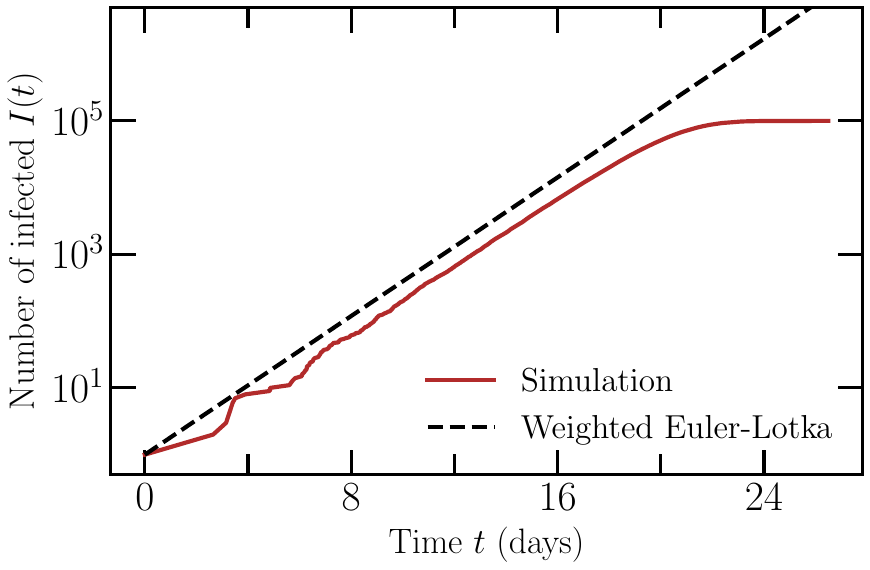}
    \caption{}
    \end{subfigure}
    \hfill
    \begin{subfigure}[b]{0.23\textwidth}
    \includegraphics[width=\linewidth]{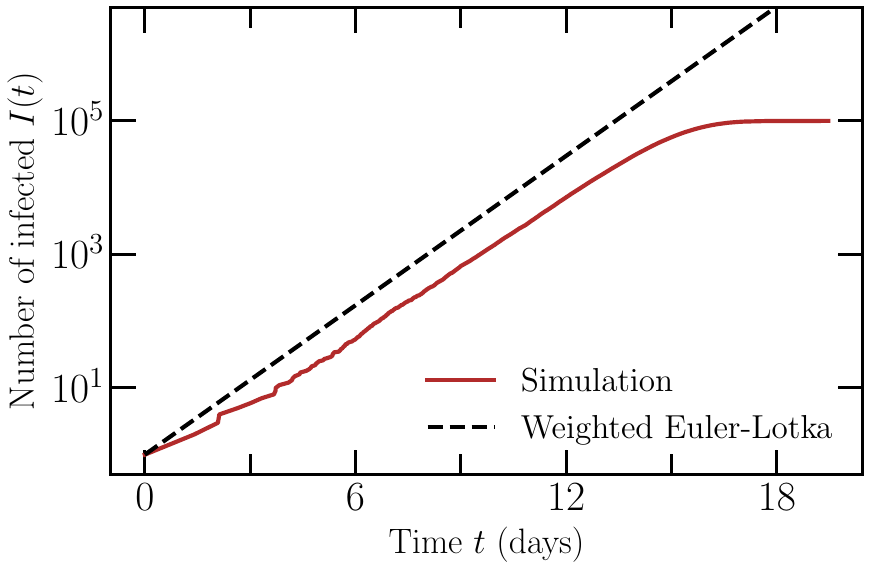}
    \caption{}
    \end{subfigure}
    \hfill
    \begin{subfigure}[b]{0.23\textwidth}
    \includegraphics[width=\linewidth]{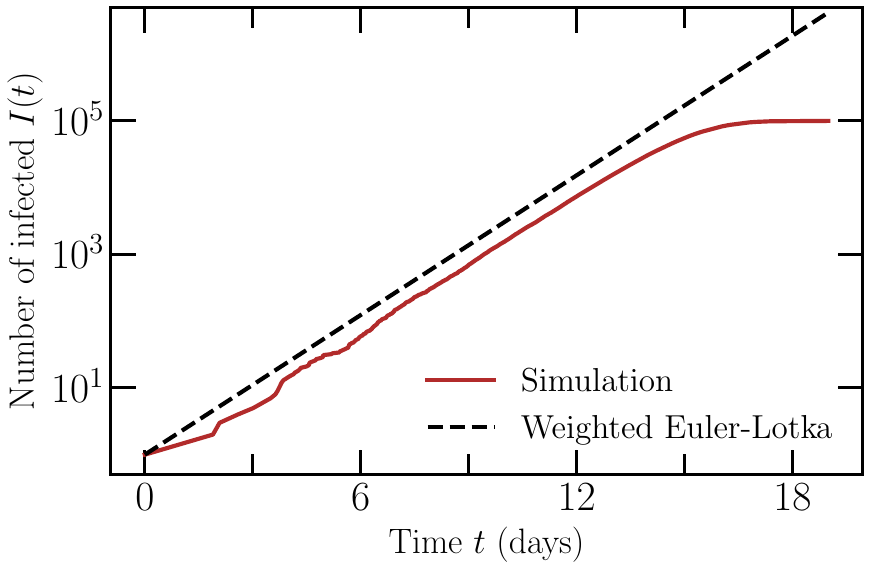}
    \caption{}
    \end{subfigure}
    \hfill
    \begin{subfigure}[b]{0.23\textwidth}
    \includegraphics[width=\linewidth]{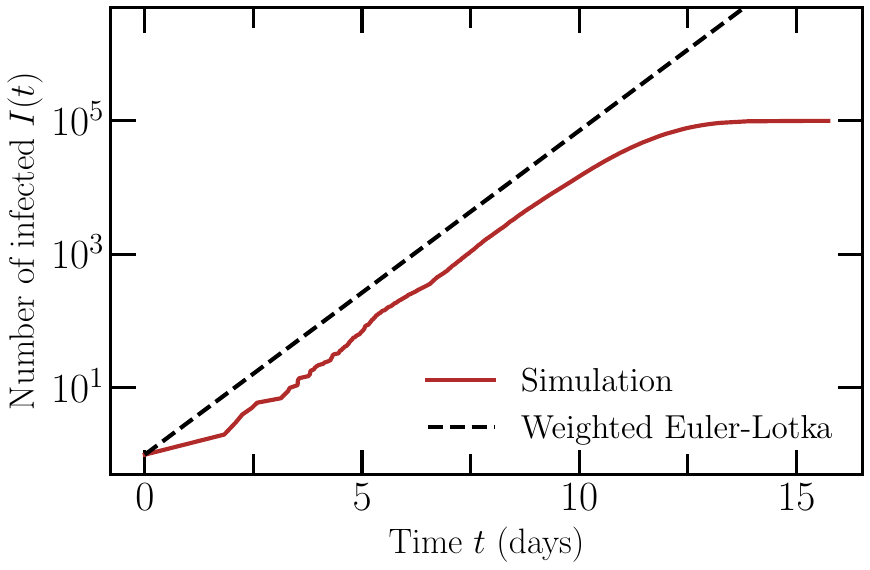}
    \caption{}
    \end{subfigure}
    \caption{Exponential spreading rate on weighted networks with two types of edges. Each network has $10^5$ nodes and two types of edges with weights $w_1$ and $w_2$. We run a simulation in which the generation time distribution for $w_i=1$ is Gamma distributed with mean $5$ and variance $ 3$. For different value of $w_i$, the generation time distribution is rescaled accordingly. (a) Edges with weight $w_1=1$ form a \WS network with parameter $k_0=6$ and $p=0.8$ and edges with weight $w_2=4$ form a \WS network with parameter $k_0=4$ and $p=0.4$.  (b) Edges with weight $w_1=1$ form a \BA network with parameter $m=3$ while edges with weight $w_2=12$ form a \WS network with parameter $k_0=6$ and $p=0.6$. (c) Edges with weight $w_1=2$ form a \ER network with mean degree $10$ while edges with weight $w_2=7$ form a \WS network with parameter $k_0=4$ and $p=0.3$.  (d) Edges with weight $w_1=1.5$ form a \ER network with mean degree $10$ while edges with weight $w_2=4.1$ form a \BA network with parameter $m=4$. The discrepancy between the solution of the weighted network Euler-Lotka equation (\ref{eq:weighted_EL}) and the numerically computed spreading rate $\Lambda_{\rm FIT}$ is of $2\%$ in (a), $6\%$ in (b), less than $1\%$ in (c), and $2\%$ in (d).}
    \label{fig:multi-type}
\end{figure}

Equation~\eqref{eq:multityp_EL} successfully predicts the exponential growth rate for weighted networks constructed as combinations of different network models, see Fig.~\ref{fig:multi-type}. In these examples, each network is constructed so that edges of each type are generated by a different network model. 

We now consider the case in which each edge $e_{ij}$ has a random weight $w_{ij}$ which is identically and independently distributed for all edges. This case is particularly instructive, as it shows more clearly the formal relation between the weighted Euler-Lotka equation and the network Euler-Lotka equation introduced in Eq.~\eqref{eq:netEL}.

\begin{figure}[!ht]
    \centering
    \begin{subfigure}[b]{0.23\textwidth}
        \centering
        \includegraphics[width=\textwidth]{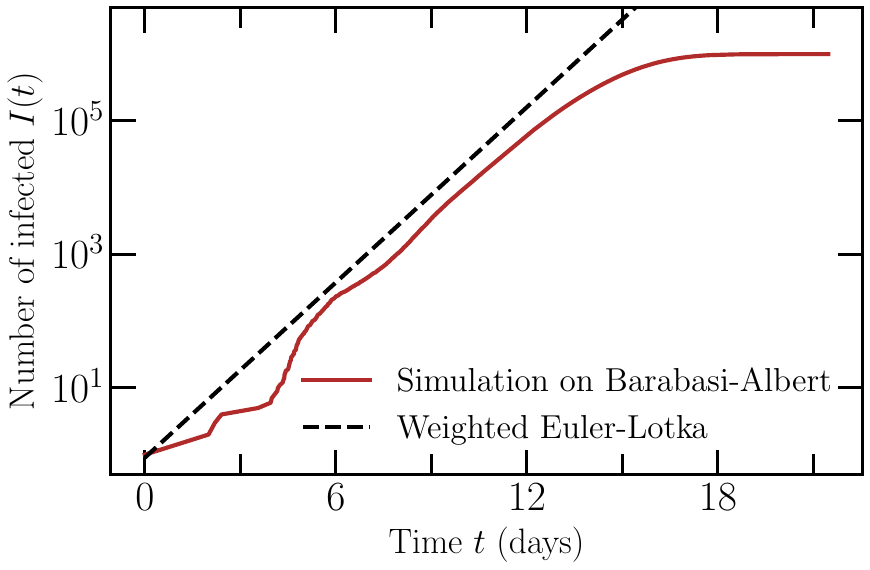}
        \caption{\BA}
        \label{fig:ba_weight}
    \end{subfigure}
    \begin{subfigure}[b]{0.23\textwidth}
        \centering
        \includegraphics[width=\textwidth]{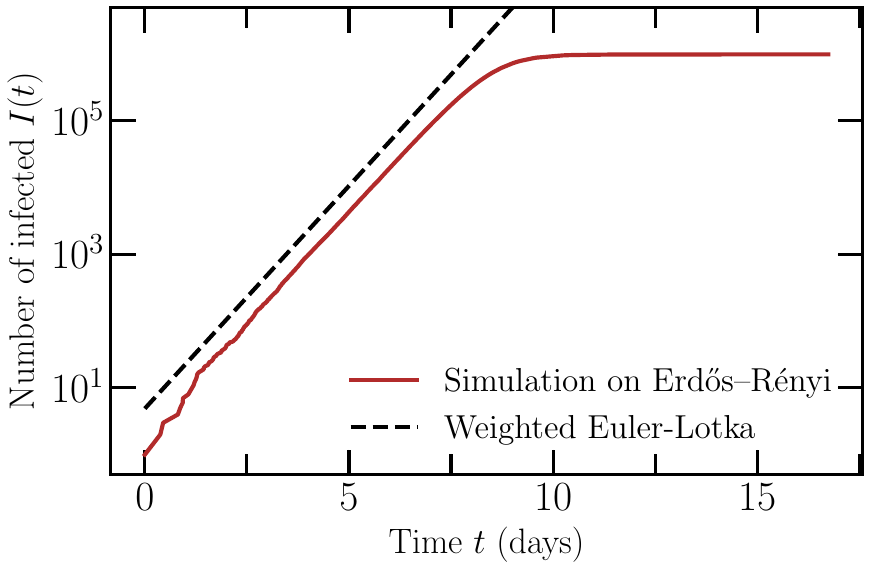}
        \caption{\ER}
        \label{fig:er_weight}
    \end{subfigure}
    \begin{subfigure}[b]{0.23\textwidth}
        \centering
        \includegraphics[width=\textwidth]{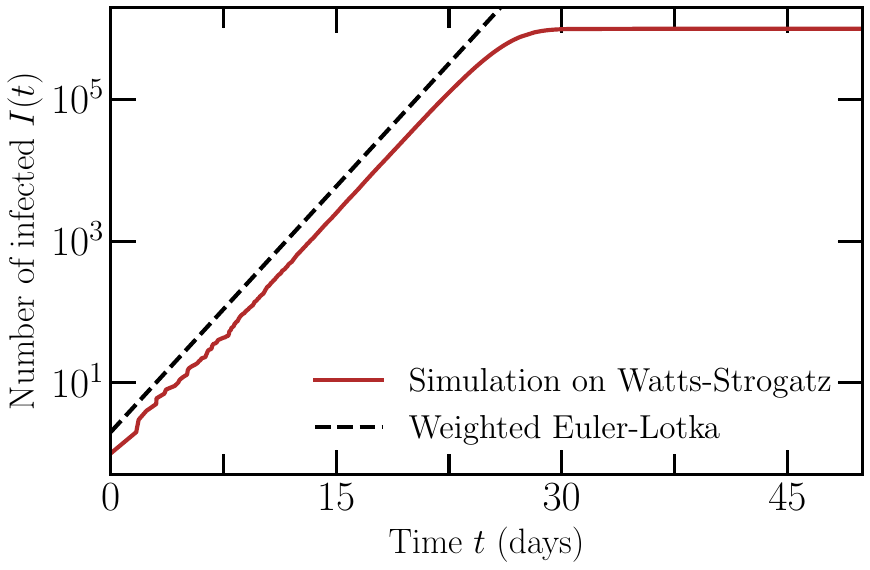}
        \caption{\WS}
        \label{fig:ws_weight}
    \end{subfigure}
    \caption{Exponential spreading rate on networks with i.i.d weights. (a): \BA network of size $10^6$ with parameter $m=7$. The infection times are lognormal with mean $5$ and variance $3$, the weights are $1$ with probability $p=0.9$ and $5$ with probability $p=0.1$.  (b): \ER networks of size $10^6$ with mean degree $5$. the infection times are Weibull distributed with shape $2$ and scale $2$. The weights follow a power-law $1/w^3$ for $w\in [1,100]$. (c): \WS network of size $10^6$ with parameter $p=0.5$ and $k=4$. The infection times are Gamma distributed with mean $3$ and variance $1$. The weights are exponentially distributed with mean $3$.  The discrepancy between the solution of the weighted network Euler-Lotka equation (Eq.~\eqref{eq:weighted_EL} and the numerically computed spreading rate $\Lambda_{\rm FIT}$ is within $9\%$ in (a), less than $1\%$ in (b), and less than $1\%$ in (c). }
    \label{fig:weight_iid}
\end{figure}

We call $q(w)\equiv Q(i|k,k')$ the probability of the weights, which in this case is independent of the degrees. Substituting into Eq.~\eqref{eq:multityp_EL} we directly obtain 
\begin{equation}
 \sum_w q(w) \left \langle e^{-\Lambda \tau}\right\rangle_w=\frac{1}{R},
 \label{eq:weighted_EL}
\end{equation}
where $\langle \dots\rangle_w$ is an average over the distribution $\psi^{(w)}(\tau)$ for a given $w$ and $R$ is the leading eigenvalue of the matrix $M_{kk'}$. Simulations for different network models and weight distributions support the predictions of Eq.~\eqref{eq:weighted_EL}, see Fig.~\ref{fig:weight_iid}.

\section{Discussion}\label{sec:discussion}
In this article, we have related the exponential spreading rate of an epidemic on a complex network with the infectiousness and the network structure. Our theory accurately predicts the spreading rate on a wide range of synthetic and real-world networks, aside from scale-free networks  where the exponential regime is hindered by finite-size effects. 
When analyzing the early stage of real epidemics, departure from an exponential spreading is often interpreted as a consequence of behavioral response or containment measures (see, e.g., \cite{maier2020}). Our results show that non-exponential epidemic trajectories are common in heavy-tailed networks, thus providing an alternative explanation for these observations.
We have focused for simplicity on static networks. Epidemic dynamics are more complex in temporal networks \cite{cai2024epidemic,ferretti2024digital}, for example, due to bursty social interactions \cite{tkachenko2021stochastic}. Our approach can be extended to these cases, for example by studying a version of Eq.~\eqref{eq:renewal} in which the matrix $M_{kk'}$ is considered to be time-dependent.

The classic Euler-Lotka equation is often used to estimate $R$ in real epidemics. However, this approach is often inaccurate, because of difficulties in estimating both $\Lambda$ and the infectiousness distribution (see, e.g., \cite{lauer2020incubation,ferretti2020quantifying,nishiura2020serial}). For this reason, approaches based on contact tracing have been developed \cite{kojaku2021effectiveness,birello2024estimates}. Our theory paves the way to use basic structural information on the interaction networks to directly estimate $R$. 

\FloatBarrier

\begin{acknowledgments}
We thank C. Castellano and I. Neri for their comments on a preliminary version of this manuscript. We thank L. Ferretti, C. Fraser, and J. Weitz for the discussions. We are grateful for the help and support provided by the Scientific Computing and Data Analysis section of Core Facilities at OIST.
\end{acknowledgments}

\appendix

\section{Perturbative expansion of the reproduction number}
\label{sec:pertub_derivation}

In this section, we derive the expression of the reproduction number $R_{\pert}$. Our strategy is to approximate the eigenvalue by first neglecting the dependence of $m_{kk'}$ on $k$. In practice, this means that we are accounting for the dependence of clustering on degree for the infecting individual, but not for the individuals receiving the infection. After this approximation, we will treat the degree correlations as a first-order perturbation of the uncorrelated case. 

Our first step is to approximate $m_{kk'}$ with its average over $k$. To do so, we use the relation $(k'-1) c_{k'}=\sum_{k} m_{k k^{\prime}} P(k| k')$, see \cite{serrano2006clustering}, which implies \begin{equation}
    m_{kk'}\approx (k'-1)c_{k'}.
    \label{eq:m_approx}
\end{equation} Using this expression, we rewrite the reproduction matrix as
\begin{equation}
    M_{kk'}=(k'-1)[1-c(k')]P(k|k')\, .
    \label{eq:weak_clustering}
\end{equation}

We recall that degree correlations are summarized by the assortativity $r$ \cite{newman2002assortative}, defined as the Pearson correlation coefficient of the degrees of connected nodes:
\begin{equation}
    r=\frac{1}{\sigma_r^2}\sum_{kk^\prime} \left[kk^\prime P(k,k^\prime)-\frac{k^2{k^\prime}^2}{\av{k}^2} P(k)P(k^\prime)\right],
    \label{assortativity_coefficient}
\end{equation}
where $\sigma_r^2=\av{k^3}/\av{k}-\left(\av{k^2}/\av{k}\right)^2$ ensures that $-1\leq r\leq 1$ and $P(k,k') = k'P(k|k')P(k')/\av{k}$ is the probability that a randomly chosen edge connects nodes of degrees $k$ and $k'$.

We now treat $r$ as a small perturbation parameter and express the leading eigenvalue using perturbation theory. To this aim, we express the reproduction matrix as
\begin{equation}
    M_{kk'}=M^{(r=0)}_{kk'}+r\delta M_{kk'},
\end{equation}
with
\begin{align}\label{eq:expand}
    M^{(r=0)}_{kk'}&=(k'-1)(1-c(k'))\frac{k}{\av{k}}P(k),\\
    \delta M_{kk'}&=(k'-1)(1-c(k'))\left(P^{(1)}(k|k')-\frac{k}{\av{k}}P(k)\right),\nonumber
\end{align}
where we have expanded $P(k|k')\approx (1-r)kP(k)/\av{k} + r P^{(1)}(k|k')$ for small $r$. The quantity $P^{(1)}(k|k')$ represents the first-order correction in $r$ to the degree distribution. We do not have an explicit expression for it, but we know that it must satisfy $\sum_{k}P^{(1)}(k|k')=1$ for the degree distribution to be normalized and $\sum_k kP^{(1)}(k|k')=k'$ from expanding $P(k,k')$ in Eq.~\eqref{assortativity_coefficient} at the first order. In Eq.~\eqref{eq:expand}, we have also assumed that the dependence of $c(k)$ on $r$ is sufficiently weak to be neglected at the first order. Strictly speaking, clustering and assortativity are not independent network properties, as assortativity constrains the maximum achievable clustering within a network \cite{serrano2006clustering}. 

In the unperturbed case $r=0$, the reproduction matrix $M_{kk'}$ has a single non-zero eigenvalue $R_\unpert$ with left and right eigenvectors $u_k^{(0)}$ and $v_k^{(0)}$, normalized such that $\sum_{k}v_k^{(0)}u_k^{(0)}=1$:
\begin{align}
    R_\unpert&=\frac{\av{k^2}-\av{k}}{\av{k}}-m_{1},\\
    u_k^{(0)} &=\frac{(k-1)(1-c_k)}{ R_\unpert}\nonumber\\
    v_k^{(0)}&=\frac{k}{\av{k}} P(k).\nonumber
\end{align}
At the order of $\mathcal{O}(r)$, the leading eigenvalue is given by
\begin{equation}
R_\pert=R_\unpert+r\sum_{kk'}u_k^{(0)} \delta M_{kk'}v_{k^\prime}^{(0)}+\mathcal{O}(r^2)\, .
\end{equation}
Using the expansion of the reproduction matrix and the unperturbed eigenvectors, we obtain:
\begin{align}
         \sum_{kk'}u_k^{(0)} \delta M_{kk'}v_{k^\prime}^{(0)}&= \frac{1}{R_\unpert} \sum_{kk'}\left[(k-1)(1-c_k) (k'-1)\right.\nonumber\\
         \times &\left.(1-c(k'))P^{(1)}(k|k')P(k')\frac{k'}{\av{k}}\right]-R_\unpert.
\end{align}

Expanding the terms involving clustering coefficients, we separate the sum in four distinct terms. The first three are 
\begin{align}
     \sum_{kk'}(k-1)(k'-1)P^{(1)}(k|k')P(k')\frac{k'}{\av{k}}&=\frac{\overline{k(k-1)^2}}{\av{k}}.
\end{align}
\begin{align}
     \sum_{kk'}c_k(k-1)(k'-1)P^{(1)}(k|k')P(k')\frac{k'}{\av{k}}&=\frac{\overline{k(k-1)^2c_k}}{\av{k}}.
     \label{eq:sym}
\end{align}
\begin{align}
     \sum_{kk'}c_{k'}(k-1)(k'-1)P^{(1)}(k|k')P(k')\frac{k'}{\av{k}}&=\frac{\overline{k(k-1)^2c_k}}{\av{k}}.
     \label{eq:antsym}
\end{align}
Equations \eqref{eq:sym} and \eqref{eq:antsym} yield the same result due to degree detailed balance, which holds because links are undirected. Indeed, we have
\begin{align}
    P(k|k')P(k')\frac{k'}{\av{k}}=P(k'|k)P(k)\frac{k}{\av{k}}.
\end{align}
Replacing $P(k|k')$ by the first-order correction in $r$ gives
\begin{align}
    \left(\frac{k}{\av{k}}P(k)(1-r)+rP^{(1)}(k|k')\right)P(k')\frac{k'}{\av{k}}=\\\left(\frac{k'}{\av{k}}P(k')(1-r)+rP^{(1)}(k'|k)\right)P(k)\frac{k}{\av{k}}.\nonumber
\end{align}
Expanding and rearranging the terms, we find
\begin{align}
   P^{(1)}(k|k')P(k')\frac{k'}{\av{k}}&=P^{(1)}(k'|k)P(k)\frac{k}{\av{k}}.
\end{align}
Therefore, detailed balance is still satisfied at the first order in $r$. Returning to the sum, the fourth term in the sum reads 
\begin{align}
     \sum_{kk'}c_kc_{k'}(k-1)(k'-1)P^{(1)}(k|k')P(k')\frac{k'}{\av{k}}.
\end{align}
To evaluate the sums, we use that $\sum_{k^{\prime}}c_{k'}(k'-1)P^{(1)}\left(k^{\prime}| k\right) = m_1$. This can be shown as follows:
\begin{align}
     m_1 &=\sum_{kk'}m_{kk'}P(k,k')\\
      &=\sum_{k^{\prime}} c_{k'}(k'-1) P\left(k^{\prime}| k\right) \\
     &=(1-r)\sum_{k^{\prime}} c_{k'}(k'-1) \frac{k^{\prime}}{\av{k}}P(k^{\prime})\nonumber \\&\quad\quad+\quad r\sum_{k^{\prime}}c_{k'}(k'-1) P^{(1)}\left(k^{\prime}| k\right)\\
    &=(1-r)m_1 +r\sum_{k^{\prime}}c_{k'}(k'-1) P^{(1)}\left(k^{\prime}| k\right)\\
    \implies & \sum_{k^{\prime}}c_{k'}(k'-1)P^{(1)}\left(k^{\prime}| k\right) = m_1,
\end{align}
where in the first step we have used the relation $(k'-1) c_{k'}=\sum_{k} m_{k k^{\prime}} P(k| k')$. Thus, we express the last term as
\begin{align}
     \sum_{kk'}c_k(k-1)c_{k'}(k'-1)P^{(1)}(k|k')P(k')\frac{k'}{\av{k}}&=m_1^2.
\end{align}
Therefore, we write the leading eigenvalue as
\begin{align}\label{eq:R_perturb}
    R_\pert &= R_\unpert(1-r)\nonumber\\
    &+ \frac{r}{R_\unpert}\left(\frac{\av{k(k-1)^2}}{\av{k}}-2m_{2}+m_1^2\right)+\mathcal{O}(r^2),
\end{align}
where we identify $m_{2}=\av{k(k-1)^2c_k}/\av{k}$ and thus recover Eq.~\eqref{eq:R_tilde}. 

We remark that $m_2$ is indirectly related to the average degree of nodes being part of a triangle. After picking randomly a triangle in the network, we pick one of the three nodes. The probability that it has degree $k$ is given by $c_k(k-1)kP(k)/(m_1\av{k})$. Thus the average degree of a node belonging to a triangle is expressed by
\begin{align}
\sum_{k}\frac{c_k(k-1)}{m_1}\frac{k^2}{\av{k}}P(k)&= \sum_{k}\frac{c_k(k-1)^2}{m_1}\frac{k}{\av{k}}P(k)\nonumber \\
&\quad+\sum_{k}\frac{c_k(k-1)}{m_1}\frac{k}{\av{k}}P(k)\nonumber\\
    &=\frac{m_2}{m_1}+1.
\end{align}
Finally, the first-order correction of the right eigenvector associated to the unique leading eigenvalue is given by
\begin{equation}
        v_k^{(1)}=v_k^{(0)}+\frac{r}{R_\unpert}\sum_{k'}\delta M_{kk'}v_{k'}^{(0)}+\mathcal{O}(r^2).
\end{equation}
Expanding the sum we obtain
\begin{align}
    \sum_{k'}\delta M_{kk'}v_{k'}^{(0)}&= \sum_{k'}(k'-1)(1-c_{k'})P^{(1)}(k|k')\frac{k^{\prime}}{\av{k}}P(k^{\prime})\nonumber \\
    -\sum_{k'}&(k'-1)(1-c_{k'})P(k^{\prime})P(k)\frac{kk^{\prime}}{\av{k}^2}\nonumber\\
    =\sum_{k'}(k'-1)&(1-c_{k'})P^{(1)}(k'|k)\frac{k}{\av{k}}P(k)-R_\unpert\frac{k}{\av{k}}P(k)\nonumber\\
    =(k-1)\frac{k}{\av{k}}&P(k)-m_1\frac{k}{\av{k}}P(k)-R_\unpert\frac{k}{\av{k}}P(k).
\end{align}
Therefore, the first-order correction to the
leading right eigenvector is given by
\begin{equation}
        v_k^{(1)}=(1-r)\frac{k}{\av{k}}P(k)+\frac{r}{R_\unpert}\left((k-1)-m_1\right)\frac{k}{\av{k}}P(k)\,.
\end{equation}
The leading eigenvector can be interpreted as the probability that an infected node is of degree $k$, at first order. In the absence of clustering and assortativity, this probability is proportional to with $kP(k)$ rather than $P(k)$ as nodes with larger degrees are more likely to receive the infection. 
In the presence of substantial positive assortativity, nodes with large degrees are even more influential, on the order of $k^2 P(k)$, as the assortativity coefficient $r$ increases the likelihood of selecting nodes with higher degrees. 
The relative difference between $R_{\pert}$ and $R_0$ as a function of the clustering coefficients and the assortativity is illustrated in Fig.~\ref{fig:phasediagram}. While positive assortativity and clustering have opposing effects on the reproduction number, the figure demonstrates that clustering exerts a stronger influence than assortativity, except in networks with heavy-tailed degree distributions, where the impact of assortativity is amplified, as we just discussed

\section{Large deviation principle}
\label{sec:LDP}
\subsection{Equation for the spreading rate}
In Sec.~\ref{sec:ELnetworks} we derived the network Euler-Lotka equation [Eq.~\eqref{eq:netEL}] from the renewal equation Eq.~\eqref{eq:renewal}. Here, we present an alternative derivation which relates the exponential spreading rate $\Lambda$ to a large deviation principle. This derivation is slightly more lengthy but will permit us to derive an approximate explicit expression for $\Lambda$ [Eq.~\eqref{eq:lambda_approx}].

In a network of infinite size, we define the exponential spreading rate as
\begin{equation}
    \Lambda = \lim_{t\to\infty}\frac{1}{t}\ln I(t)\, ,
    \label{def_Lambda}
\end{equation}
provided that this limit exists. 
We first express the number of infected $I(t)$ in terms of lineages. The epidemic starts with a single infected individual, which we call the root node, and the epidemic progresses as infected individuals spread the infection to their neighbors. We call a leaf a node that is infected but has not yet transmitted the disease to any of its neighbors. We define $L(n,t)$ as the number of leaves at time $t$ that are at a distance $n$ from the root. The epidemic passed through $n$ different nodes before reaching those leaves. The total number of leaves is always smaller or equal to the total number of infected, but grows exponentially at the same rate, so that
\begin{equation}
    \Lambda = \lim_{t\to \infty}\frac{1}{t}\ln \sum_n L(n,t).  \label{lambda_wrt_lineages}
\end{equation} 
We express the total number of leaves in the limit $t\to\infty$ as
\begin{equation}
    \sum_n L(n,t)\approx \sum_n Z_n p_t(n),
\end{equation}
where $Z_n$ is the average number of nodes at a distance $n$ from the root and $p_t(n)$ is the probability that on a given lineage the infection has spread to a distance $n$ by time $t$. This approximation rests on the assumption that lineages are independent, which is effectively the case for large time. Therefore, we obtain
\begin{equation}
    \Lambda = \lim_{t\to \infty}\frac{1}{t}\ln \sum_n Z_n p_t(n).
\end{equation}
This representation allows us to separate the properties of the infection times and of the network: $Z_n$ only depends on the structure of the network while $p_t(n)$ only depends on $\psi(\tau)$. In the case where $p_\psi=1$, then $Z_n$ can be interpreted as the number of newly infected at generation $n$ and scales as $Z_n\sim R^n$, where $R$ is the reproduction number. Therefore, we have
\begin{equation}
    \Lambda= \lim_{t\to \infty}\frac{1}{t}\ln \left\langle R^n\right\rangle_n.\label{Lambda_wrt_Zn}
\end{equation} 
where $\left\langle \dots \right\rangle_n$ represents the average over the distribution $p_t(n)$. In this representation, the variable $n$ depends on $t$ and can be interpreted as a counting process $n(t)$ with generation-time $\psi(\tau)$, such that $p_t(n)$ is the probability that a lineage is of size $n$ at time $t$.

We now follow the same steps as in Ref.~\cite{pigolotti2021generalized}. We define the variable $\omega=n/t$, so that $\omega$ tends to a constant as $t$ grows large. We say that $\omega$ satisfies a large deviation principle if
\begin{equation}
    p(\omega) \asymp e^{-t I^{(\omega)}(\omega)},
\end{equation}
where $I^{(\omega)}(\omega)$ is called the rate function and where the sign $\asymp$ means that  as $t\to\infty$ the dominant part of $p(\omega)$ is the decaying exponential $e^{-t I^{(\omega)}(\omega)}$. Assuming that $\omega$ satisfies a large deviation principle and that the rate function is convex, we apply the G\"{a}rtner-Ellis theorem 
\begin{equation}
    \Psi^{(\omega)}(q)=\sup_\omega\left[q \omega-I^{(\omega)}(\omega)\right]\, ,
    \label{involution_omega}
\end{equation}
where $\Psi^{(\omega)}(q)=\lim _{t \rightarrow \infty} t^{-1} \ln \langle e^{q t \omega}\rangle_\omega$ is the the scaled cumulant generating function of $\omega$. Using in succession Eqs.~\eqref{Lambda_wrt_Zn} and ~\eqref{involution_omega} we obtain
\begin{equation}
\Lambda=\sup_\omega\left[\omega \ln R-I^{(\omega)}(\omega)\right].
\label{lambda_first_deviation_relation}
\end{equation}
To determine $I^{(\omega)}(\omega)$, we remember that $n(t)$ is described by a renewal process with generation time $\psi(\tau)$. This means that its inverse $t(n)$ is the sum of $n$ i.i.d. random variables distributed according to $\psi(\tau)$. Following Ref.~\cite{glynn1994large}, their rate functions are related by 
\begin{equation}
    I^{(\omega)}(x)=x I^{(t)}(1 / x).
    \label{inverse_relation}
\end{equation}
The scaled cumulant generating function of $t$ can be expressed in terms of the probability density distribution of the infection times:
\begin{align}
    \Psi^{(t)}(q)&=\lim _{n \rightarrow \infty} \frac{1}{n} \ln \left\langle\mathrm{e}^{q \sum_{i=1}^n \tau_i}\right\rangle_\tau\nonumber\\
    &=\lim _{n \rightarrow \infty} \frac{1}{n} \ln \prod_{i=1}^n\left\langle\mathrm{e}^{q \tau_i}\right\rangle_\tau\nonumber\\
    &= \ln\left\langle e^{q \tau}\right\rangle_\tau.
\end{align}
Applying once again the G\"{a}rtner-Ellis theorem, we find
\begin{align}
    I^{(t)}(1/x)&=\sup _q\left[\frac{q}{x}-\ln\left\langle e^{q \tau}\right\rangle_\tau\right]\, .
\end{align}
Combining this result with Eq.~\eqref{inverse_relation}, we obtain 
\begin{align}
    I^{(\omega)}(x)&=x \sup _q\left[\frac{q}{x}-\ln\left\langle e^{q \tau}\right\rangle_\tau\right].
\end{align}
Substituting this in Eq.~\eqref{lambda_first_deviation_relation}, we find
\begin{align}\Lambda&=\sup_\omega\inf_q\left[\omega \ln R-q+\omega\ln\left\langle e^{q \tau}\right\rangle_\tau\right]
    \label{lambda_second_deviation}
\end{align}
The extremality condition with respect to $\omega$ is expressed by
\begin{equation}
    \left\langle e^{q \tau}\right\rangle_\tau=\frac{1}{R}
\end{equation}
Substituting it back in Eq.~\eqref{lambda_second_deviation}, we obtain $\Lambda = -q^\text{inf}$. Inserting this in the extremality condition gives the Euler-Lotka equation, Eq.~\eqref{eq:netEL}.

\subsection{Gaussian approximation of the spreading rate}
\label{gaussian}
In this section, we derive the approximate expression for $\Lambda$, Eq.~\eqref{eq:lambda_approx}. In the limit $t\to\infty$, the distribution of the counting process $n(t)$ tends to a Gaussian with mean $\mu_n=t/\langle\tau\rangle$ and variance $\sigma_n^2= t \sigma^2/\langle\tau\rangle^3$ where $\sigma ^2 = \langle\tau^2\rangle-\langle\tau\rangle^2$ is the variance of the generation-time distribution $\psi(\tau)$. Using Eq.~\eqref{Lambda_wrt_Zn}, we thus represent $\Lambda$ as an average over a Gaussian distribution:
\begin{align}
\Lambda&= \lim_{t\to \infty}\frac{1}{t}\ln \left \langle e^{n\ln R}\right\rangle_n\\
&\approx \lim_{t\to \infty}\frac{1}{t}\ln \int_{-\infty}^{+\infty} e^{-\frac{(n-\mu_n )^2}{2 \sigma_n ^2}+n\ln R}\mathrm{d}n\\
&= \lim_{t\to \infty}\frac{1}{t}\ln \left( 
e^{\mu_n \ln R +((\ln R)^2 \sigma_n^2)/2}
\right)\\
&=\frac{\ln R}{ \left \langle\tau\right\rangle}+(\ln R)^2\frac{\sigma^2}{2 \left \langle\tau\right\rangle^3}\, .
\end{align}
This formula illustrates how a larger variance contributes to a faster spreading epidemic, while a larger mean tends to slow down the spreading rate.

\section{Simulations}
\label{sec:simulations}
\subsubsection*{Generating networks}
We generate Erd\H{o}s-R\'enyi, Barab\'asi-Albert, and Watts–Strogatz networks using standard algorithms \cite{posfai2016network}. We also generate networks with arbitrary degree distribution by means of the configuration model \cite{posfai2016network} which assigns  a number of stubs (half-links) to every node according to a given degree sequence. Then each pair of stubs is randomly chosen and connected via a link. Self-links and multiple links can appear, but their number is of the order $\mathcal{O}(1/N)$, where $N$ is the number of nodes in the network. In the limit of large $N$, the network possesses a tree-like structure (no short loops) and the degrees of connected nodes are uncorrelated ($r\approx 0$). 

We tune the assortativity in a given network as follows. We choose two pairs of connected nodes $a,b,c,d$, arranged such that $k_a\leq k_b \leq k_c \leq k_d$. We rewire the links by forming new pairs $a-b$ and $c-d$ to increase assortativity, or $a-d$ and $b-d$ to decrease assortativity. This operation preserves the degree distribution while inducing degree correlation.

To generate networks with a tunable clustering coefficient, we implement an algorithm that generalizes the configuration model \cite{serrano2005tuning}. In our implementation, we start with a degree sequence generated from a distribution $P(k)$ and choose an exponent $\alpha$ so that the number of triangles being part of a node of degree $k$ is on average $(k-1)c_k$, so that $c_k\propto (k-1)^{-\alpha}$. See Ref.~\cite{serrano2005tuning} for more details.

We collected real networks available from various databases: SNAP \cite{snapnets}, ICON \cite{clauset2016colorado}, and KONECT \cite{konect}. We selected networks of size $N>1000$, only possessing undirected links, not temporal, and not bipartite. The  properties of the selected networks and their estimated reproduction number are listed in \cite{suppl}.

\subsubsection*{Simulations}
We simulate the average trajectory of an epidemic over $500$ trajectories for every network. We repeat the simulations for three different infection time distributions: a Gamma distribution with mean $7$ and variance $1$:
\begin{equation}\label{eq:pdf_gam}
    \psi_{\rm gam}(\tau)=\frac{e^{-\frac{m \tau }{v}} \left(\frac{v}{m \tau
   }\right)^{-\frac{m^2}{v}}}{\tau  \Gamma
   \left(\frac{m^2}{v}\right)},\quad m=7,v=1\, ;
\end{equation}
a Log-normal distribution with mean $5$ and variance $3$
\begin{equation}\label{eq:pdf_log}
    \psi_{\rm log}(\tau)=\frac{\exp \left(-\frac{\left(\log (t)-\log
   \left(\frac{m^2}{\sqrt{m^2+v}}\right)\right)^2}{2 \log
   \left(\frac{v}{m^2}+1\right)}\right)}{\sqrt{2 \pi } t
   \sqrt{\log \left(\frac{v}{m^2}+1\right)}},\quad m=5,v=3\, ;
\end{equation}
and a Weibull distribution with scale $2$ and shape $2$
\begin{equation}\label{eq:pdf_wei}
    \psi_{\rm wei}(\tau)=\frac{1}{2} e^{-\frac{\tau ^2}{4}} \tau.
\end{equation}
The epidemic starts with one infected individual, chosen uniformly at random. To simulate the epidemics, we implemented a next-reaction scheme \cite{gibson2000efficient}. A detailed description of our simulation algorithm and its related code can be found in Ref.~\cite{cure2024fast}.

\section{Power-law fit of the trajectories}
\label{power-law-fit}
\begin{figure}[!h]
    \centering
    \begin{subfigure}{0.15\textwidth}
        \centering
        \includegraphics[width=\linewidth]{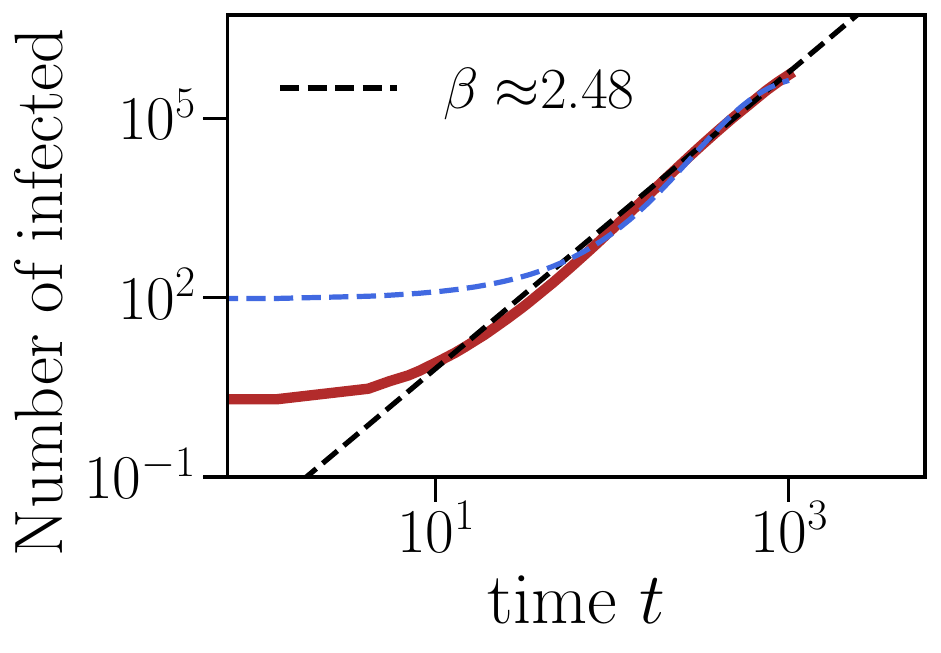}
        \caption{California}
    \end{subfigure}%
    \hfill
    \begin{subfigure}{0.15\textwidth}
        \centering
        \includegraphics[width=\linewidth]{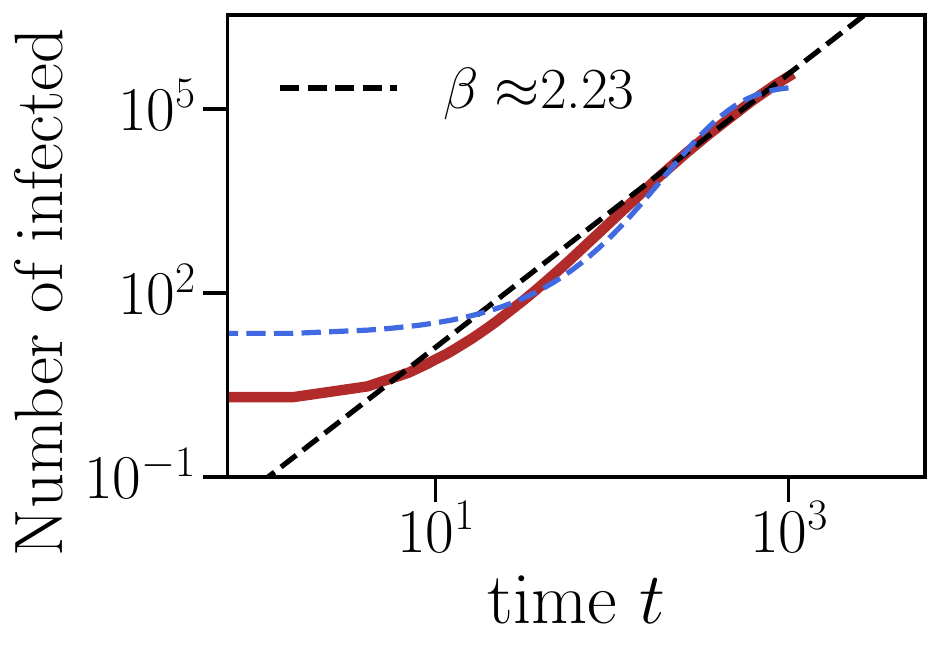}
        \caption{Pennsylvania}
    \end{subfigure}
    \hfill
    \begin{subfigure}{0.15\textwidth}
        \centering
        \includegraphics[width=\linewidth]{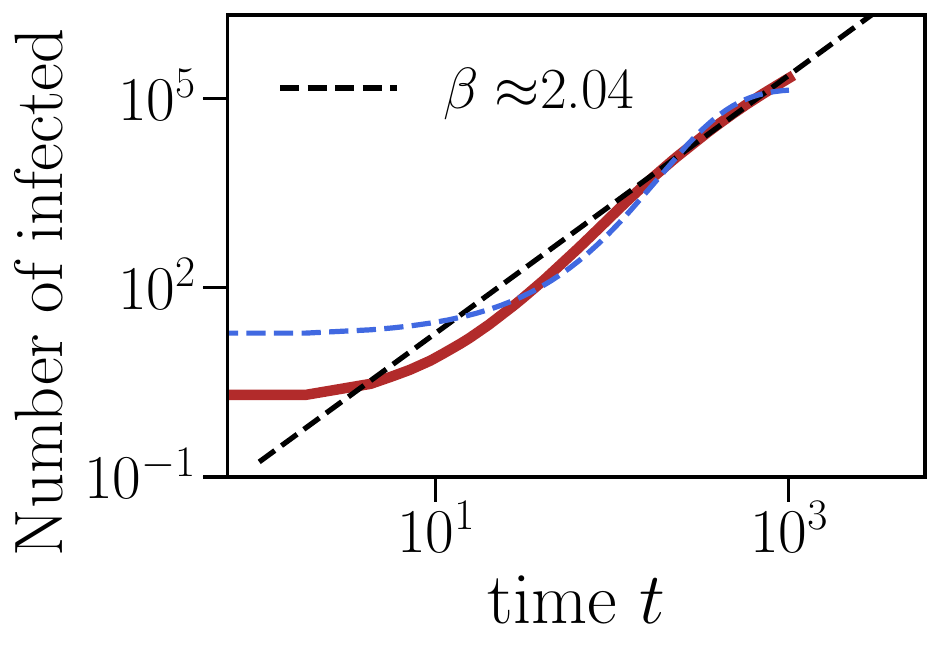}
        \caption{Texas}
    \end{subfigure}\hfill
    \begin{subfigure}{0.15\textwidth}
        \centering
        \includegraphics[width=\linewidth]{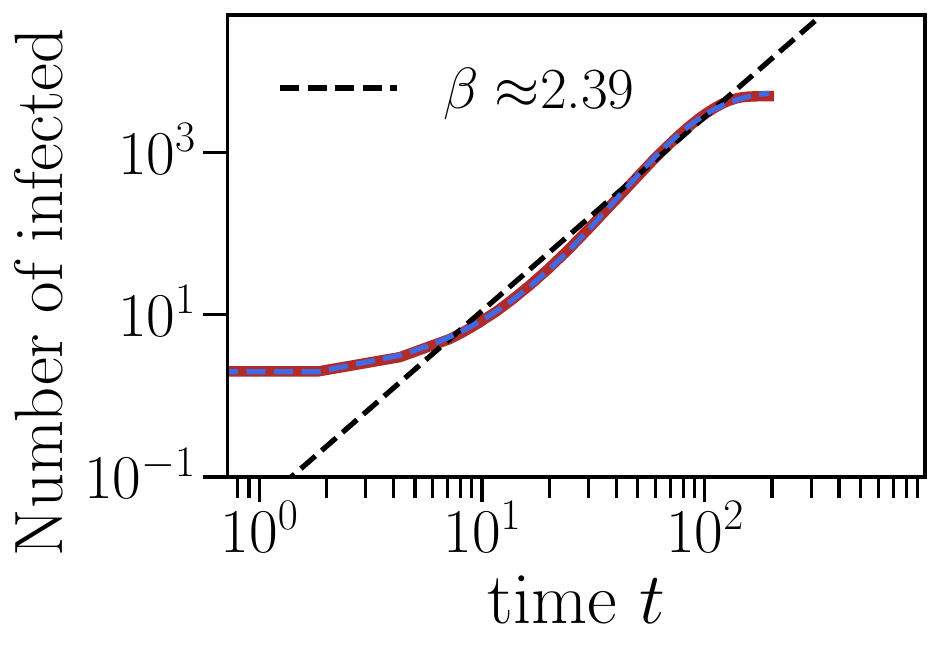}
        \caption{US powergrid}
    \end{subfigure}
    \begin{subfigure}{0.15\textwidth}
            \centering
            \includegraphics[width=\linewidth]{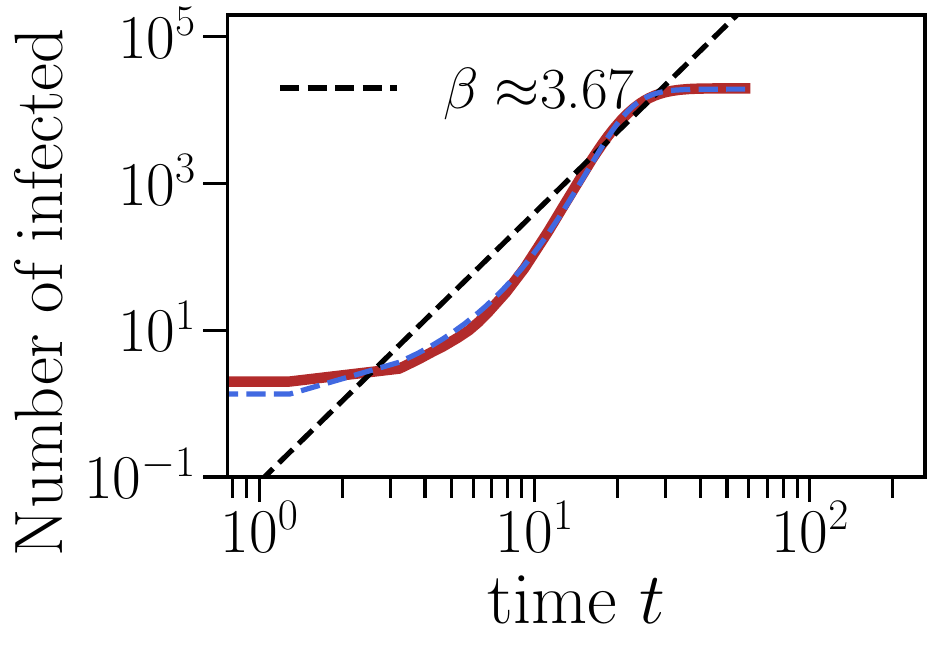}
            \caption{Citation}
    \end{subfigure}\hfill
    \caption{Comparison between power-law fit and generalized logistic fit for road networks. We plot the average epidemic trajectories (red dots) for the three road networks (panels (a)-(b)-(c)), the US powergrid (d) and the condensed matter citation network (e). We fit a power-law (blue) and the generalized logistic function (black) for each trajectory. The infection times are lognormally distributed with mean $5$ and variance $3$.}
    \label{fig:powerlaw_fit}
\end{figure}

The fitting procedure of a generalized logistic function described in Methods fails for epidemics on the California, Pennsylvania, and Texas road networks \cite{snapnets}. Instead, their behavior appears more similar to a power-law $I(t)\propto t^\beta$. We therefore fit a power-law to three U.S. road networks, the U.S. power grid, and compare it to the Condensed Matter citation network, that has a clear exponential phase, see Fig. \ref{fig:powerlaw_fit}. The power-law fits are excellent for the US road networks, while the generalized logistic function provides a better fit for the U.S. power grid network. This supports that finite-size effects ($\eta=0.497$) are more likely to explain the deviation from the predicted spreading rate for the power grid network ($\Lambda_{\rm EL}= 0.196$ versus $\Lambda_{\rm FIT}=  0.351$) rather than the presence of a power-law behavior.

\section{Comparison between estimates of the reproduction number}
\label{sec:R_vs_R0}
It is possible, in principle, to use $R_0$ instead of $R_\pert$ to predict the spreading rate by means of the network Euler-Lotka equation. Here, we test the performance of this simpler approach.

\begin{figure}[!h]
    \centering
    \begin{subfigure}[b]{0.23\textwidth}
        \centering
        \includegraphics[width=\textwidth]{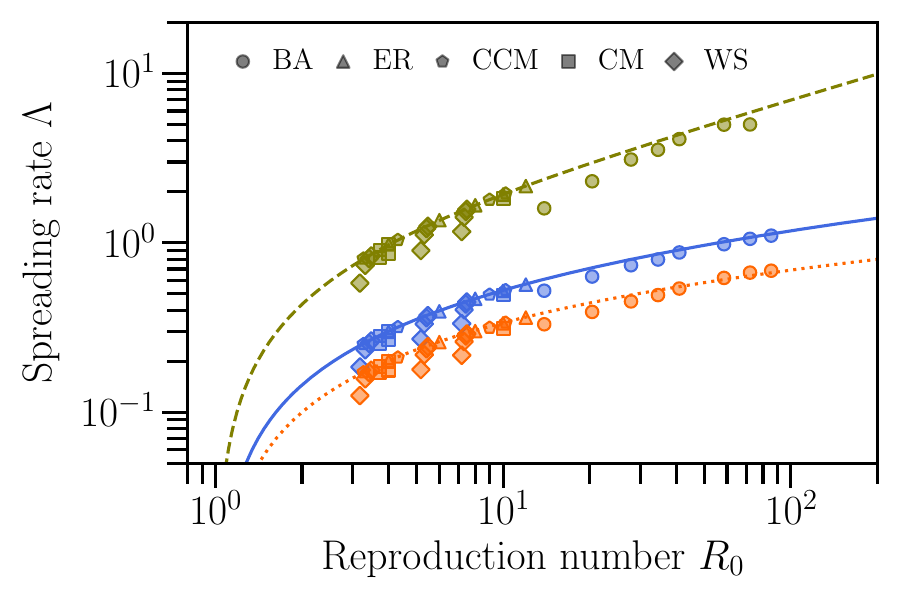}
        \caption{}
    \end{subfigure}
    \begin{subfigure}[b]{0.23\textwidth}
        \centering
        \includegraphics[width=\textwidth]{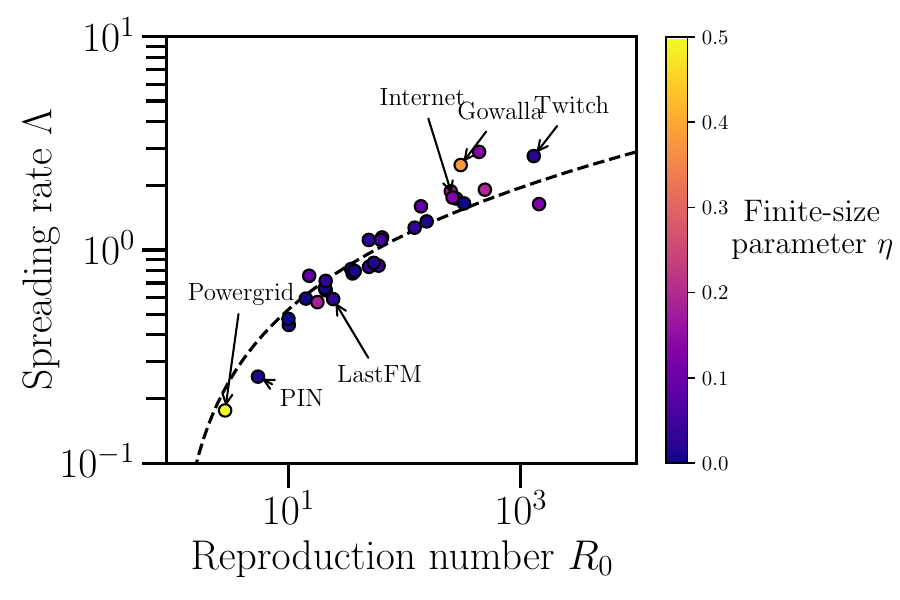}
        \caption{}
    \end{subfigure}
    \caption{ Prediction of the spreading rate using $R_0$ instead of $R_{\rm pert}$ as the reproduction number.  (a) Analogous to Fig.~\ref{fig:2}c. (b) Analogous to Fig.~\ref{fig:3}c. 
}
    \label{fig:comparison_r0_vs_rpert}
\end{figure}

For synthetic networks, the predictions of $\Lambda$ by the network Euler-Lotka equation using $R_\text{pert}$ as the reproduction number are very accurate [Fig.~\ref{fig:comparison_r0_vs_rpert}(a)].  
In contrast, using $R_0$ leads to an overestimate of the spreading rates [Fig.~\ref{fig:comparison_r0_vs_rpert}(b)].
The reason is that $R_0$ 
tends to over-estimate the reproduction number (see Table~1 in the Main Text), most noticeably for \BA and \WS networks.  
The relative errors on $\Lambda$ using 
$R_\text{pert}$ and $R_0$ for different synthetic networks and generation time distributions are listed in the Tables in the Supplementary Material \cite{suppl}.

\begin{figure}[!h]
    \centering
    \begin{subfigure}[b]{0.23\textwidth}
        \centering
        \includegraphics[width=\textwidth]{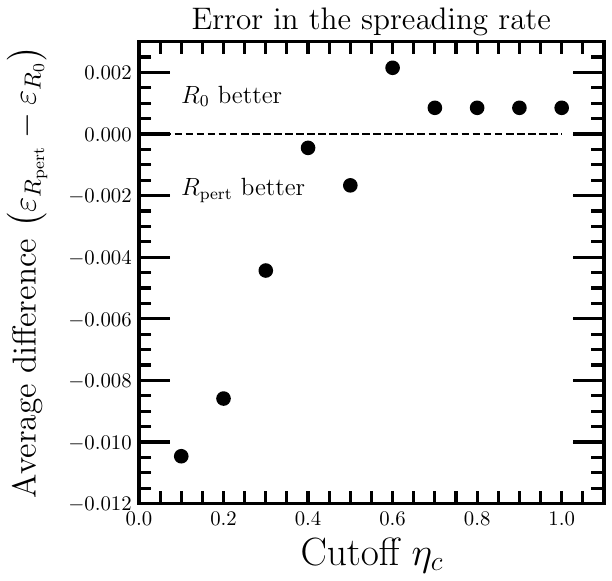}
        \caption{}
    \end{subfigure}
    \begin{subfigure}[b]{0.23\textwidth}
        \centering
        \includegraphics[width=\textwidth]{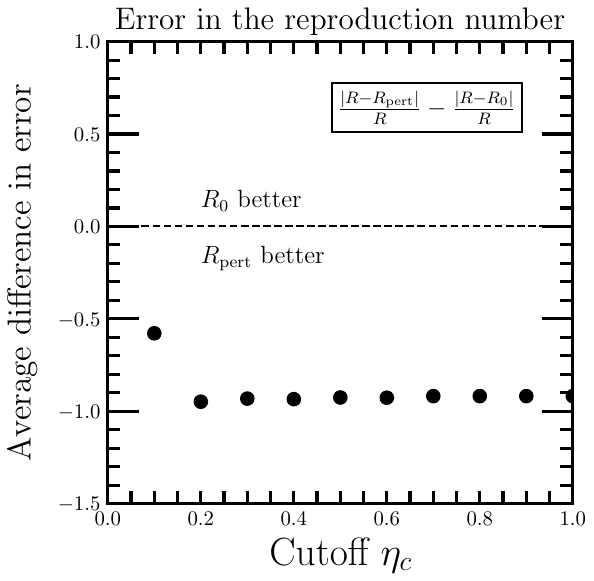}
        \caption{}
    \end{subfigure}
    \caption{ Performance of $R_\text{pert}$ and $R_0$ for different values of the finite-size parameter $\eta$. (a) Mean difference between the relative errors of spreading rates estimated using $R_\text{pert}$ and $R_0$ as a function of the finite-size parameter $\eta$. Each dot represents the average of $\varepsilon_{R_\pert}-\varepsilon_{R_0}=2|\Lambda_{R_\text{pert}} - \Lambda_\text{fit}| / (\Lambda_{R_\text{pert}} + \Lambda_\text{fit}) - 2|\Lambda_{R_0} - \Lambda_\text{fit}| / (\Lambda_{R_0} + \Lambda_\text{fit})$ across all cases where $\eta \leq \eta_c$. (b) Mean difference between the errors of $R_\text{pert}$ and $R_0$ as estimators of the reproduction number $R$ is a function of the finite-size parameter $\eta_c$. Each dot represents the average of $|R_\text{pert} - R| / R - |R_0 - R| / R$ across all cases where $\eta \leq \eta_c$.}
    \label{fig:Rpert_vs_eta}
\end{figure}

\begin{figure}
    \centering
    \begin{subfigure}[b]{0.15\textwidth}
        \centering
        \includegraphics[width=\textwidth]{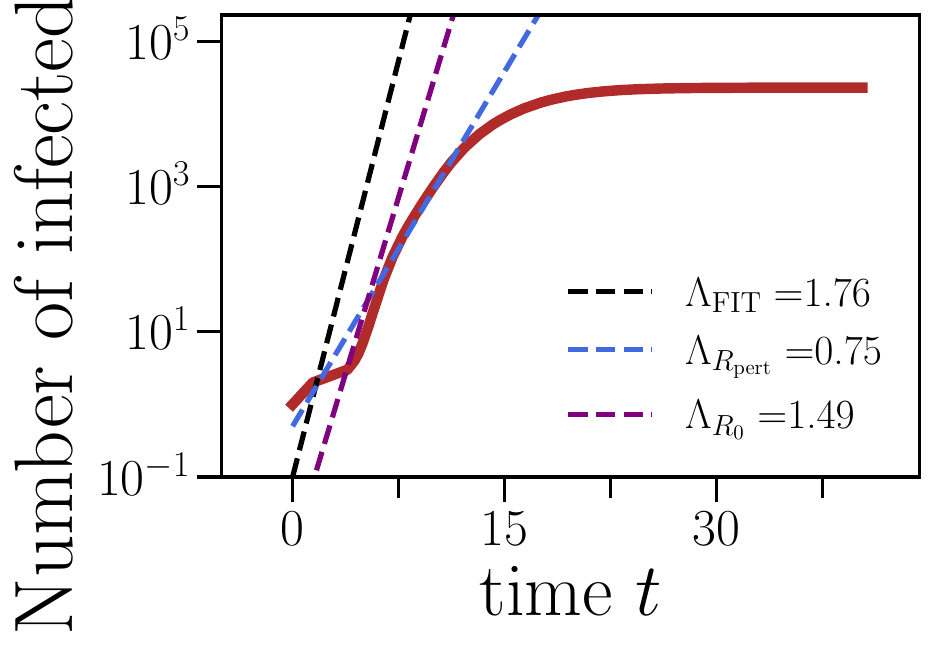}
        \caption{}
    \end{subfigure}
    \begin{subfigure}[b]{0.15\textwidth}
        \centering
        \includegraphics[width=\textwidth]{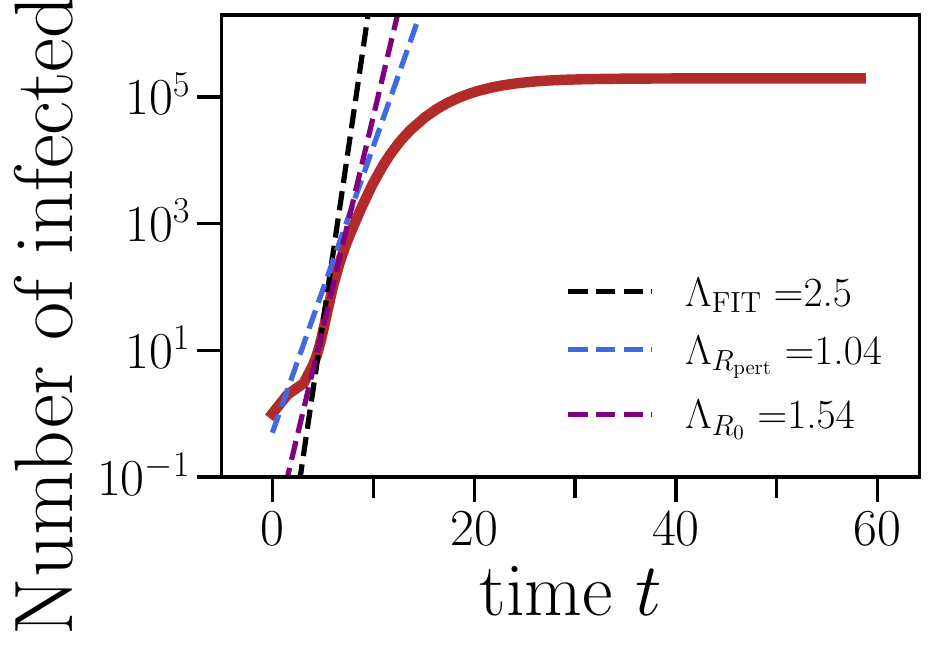}
        \caption{}
    \end{subfigure}
    \begin{subfigure}[b]{0.15\textwidth}
        \centering
        \includegraphics[width=\textwidth]{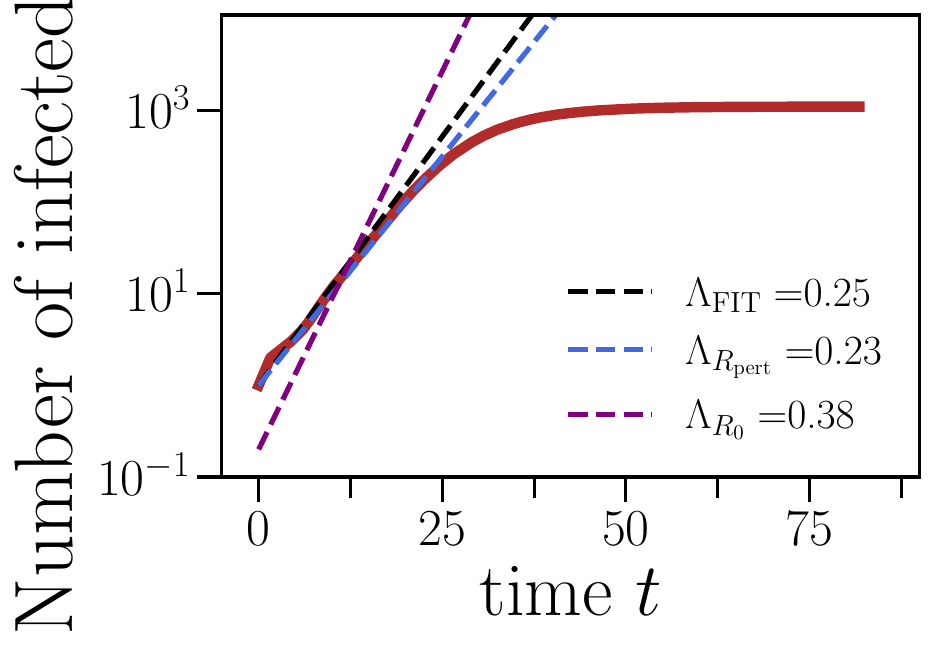}
        \caption{}
    \end{subfigure}
    \caption{ Illustration of the difference between $R_\pert$ and $R_0$ for various networks. We simulate the average epidemic using a lognormal distribution on three different networks: (a) internet, (b) gowalla and (c) PIN. We compare the difference in the spreading rates using different estimates of the reproduction number. The first two networks do not exhibit a clear exponential regime (internet: $\eta=0.13$, Gowalla: $\eta=0.38$, opposed to PIN: $\eta=0.01$)}
    \label{fig:examples-R0}
\end{figure}

For real-world networks, the predictions of $\Lambda$ using $R_\text{pert}$ are more accurate on average than those using $R_0$, as long as $\eta<0.5$ [see Fig.~\ref{fig:Rpert_vs_eta}(a)]. For larger values of $\eta$, $R_0$ provides slightly more accurate predictions of the spreading rate [Fig.~\ref{fig:Rpert_vs_eta}(a)], although $R_\text{pert}$ still provides a better estimate of the reproduction number [Fig.~\ref{fig:Rpert_vs_eta}(b)].  As discussed in the main text, this suggests a compensatory effect acting for large $\eta$.  Examples of comparison between the exponential spreading rates estimated using $R_\pert$ and $R_0$ are shown in Fig. ~\ref{fig:examples-R0}

\FloatBarrier
\bibliography{references}

\end{document}